%% file: main.tex
{\global\def\bigPage{
       \setlength{\topmargin}{-0.5in}
        \setlength{\textheight}{9in}
        \setlength{\oddsidemargin}{+0.0in}
        \setlength{\textwidth}{6.5in}
        }
}
\documentclass[11pt]{article}
\usepackage{setspace}
\usepackage{soul}
\usepackage{float}
\usepackage{graphicx}
\graphicspath{{figures/}}
\usepackage{amsmath,amssymb,dsfont,verbatim}
\usepackage{array}
\usepackage{import}
\usepackage{subcaption}
\usepackage{url, hyperref}
\usepackage{cite}
\usepackage{xcolor}
\usepackage{booktabs} 
\usepackage{tikz}
\usepackage{paralist}
\subimport{custom/}{operators.tex}

\subimport{custom/}{environments.tex}

\usepackage{etoolbox}
\newtoggle{short}
\toggletrue{short}
\togglefalse{short}

\let\OLDthebibliography\thebibliography
\renewcommand\thebibliography[1]{
  \OLDthebibliography{#1}
  \setlength{\parskip}{0pt}
  \setlength{\itemsep}{0pt plus 0.3ex}
}

\bigPage
\begin{document}%
\title{\vspace{-.75in}Networked Signal and Information Processing\vspace{-.25in}}%
\author{Stefan Vlaski, Soummya Kar, Ali H. Sayed, and Jos{\'e} M.~F.~Moura}%
\date{\vspace{-5ex}}
\maketitle
\abstract{\noindent \textbf{Abstract.} The article reviews significant advances in networked signal and information processing, which have enabled in the last 25 years extending decision making and inference, optimization, control, and learning to the increasingly ubiquitous environments of distributed agents. As these interacting agents cooperate, new collective behaviors emerge from local decisions and actions. Moreover, and significantly, theory and applications show that networked agents, through cooperation and sharing, are able to match the performance of cloud or federated solutions, while \textcolor{black}{offering the potential for improved privacy}, increasing resilience, and saving resources.\iftoggle{short}{\footnote{A longer version of this manuscript, with examples and illustrative applications, is available on arXiv:2210.13767.}}{}}

\vspace{-4mm}\section{Introduction}
\begin{figure}[htb]
  \centering \includegraphics[width=.9\columnwidth]{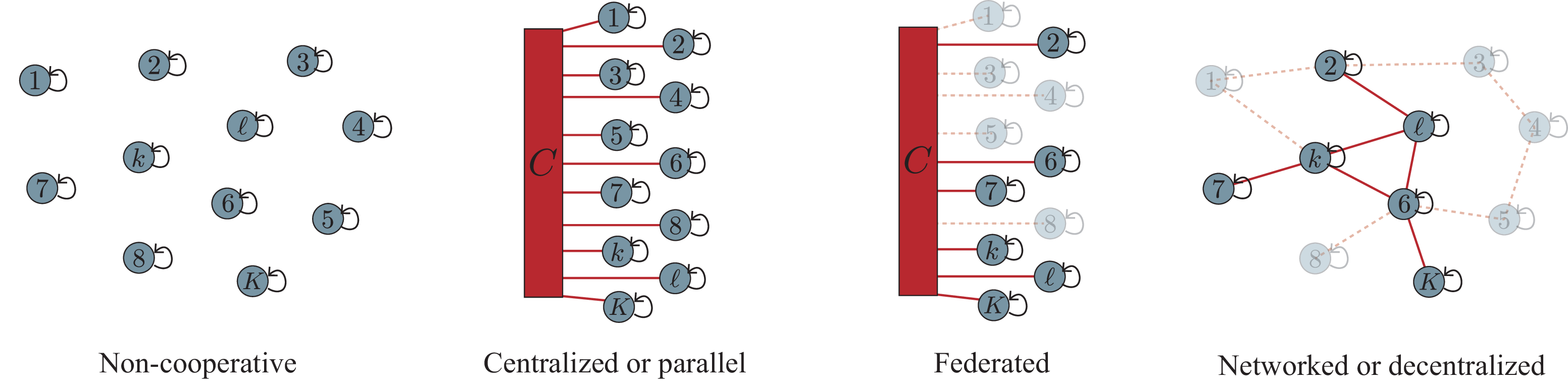}
\caption{\small Taxonomy of networked multi-agent systems.}\label{fig:taxonomy}
\end{figure}

\vspace{-2mm}\label{sec:intro}
Since its beginnings, throughout the past century and still dominant at the turn of the 21st century, the signal and information processing (SIP) prevailing paradigm has been to process signals and information by stand-alone systems or central computing units, with no cooperation or interaction among them, see left of Fig.~\ref{fig:taxonomy}. This focus has led to tremendous progress in a wide range of problems in speech and image processing, control and guidance, estimation and filtering theories, communications theory, and array processing, with enormous impact in telecommunication and wireless, audio, medical imaging, multimedia, radar, and other application areas. In the nearly 25 years since the turn of the century, each of these areas has progressed rapidly, in large part due to increases in computational resources along with the availability of data, giving rise to a variety of advanced data-driven processing tools. At the end of the century, we also witnessed significant technological progress from massive layouts of fiber at the backbone, to successes in high speed wireless and wifi deployments, to chip advances combining in a single miniature inexpensive platform functionalities like sensing, storage, communication, and computing, and to breakthroughs in network protocols and software. This progress has led for example to the launching of hundreds, soon thousands and millions, of inexpensive sensing devices (we will call here \textit{agents}) that sense, compute, communicate and are networked, ushering a paradigm shift in SIP. Initially, the agents observed data independently of one another and simply forwarded their raw data to the cloud, with no local processing in a \textit{centralized} architecture, see  Fig.~\ref{fig:taxonomy}.
 \textit{Parallel} architectures soon emerged where agents started processing their local data, transferring only their (local) inferences to a fusion center. The fusion center aggregated the locally processed data and orchestrated the computations that occurred in parallel at the individual agents. While traditionally computation and communication occurred in a synchronous fashion, synchrony requirements were relaxed, like with \textit{federated} learning architectures, third from left in Fig.~\ref{fig:taxonomy}. But as a result of scenarios with abundant data available at dispersed networked locations, such as sensor networks that monitor large geographical regions, or robotic swarms that collaborate over a common task, or social networks of many interacting agents, a new critical trend started to materialize. This led to decentralization and democratization of technology and, towards the middle and end of the first decade of this century, signal and information processing  moved definitely from parallel, federated, or edge architectures,\footnote{We interpret an edge architecture as a layered or hierarchical federated architecture.} to a \textit{distributed}, \textit{decentralized}, or \textit{networked} paradigm. The agents sense and process their own data and then cooperate with other agents. They transmit to and exchange information with agents in their vicinity. It marked the appearance of \textit{networked} elementary processing units, with each unit collecting and processing data and sharing their information with immediate neighbors. Individual agents are now capable of local inference decisions and limited  actions. The coupling among the agents gives rise to powerful network structures that open up vast opportunities for the solution of more complex problems by tapping into the power of the group. Examples of such networked systems are plentiful, including instrumented critical infrastructures like water, gas, financial networks, smart grids, as well as networked mobile devices, swarms of drones, autonomous vehicles, or populations of individuals. The interconnectedness of the agents within the network allows for their cooperation to rise from local to global coherent decision and action. To study, understand, and steer the occurrence of these global behaviors adds new layers of complexity. More advanced analytical tools became necessary to combine local processing with cooperation among the agents. This ushered the design of new processing algorithms, new methods to derive performance guarantees and assess their quality, to examine the effect of agents coupling on network stability, to endow agents with adaptation and learning abilities and with the capacity to handle privacy, and to enable such networks to contribute to multiple tasks at once. \emph{Distributed}\textcolor{black}{, \emph{decentralized},} or \textit{networked} architectures achieve aggregation and coordination through device-to-device or peer-to-peer interactions. Computation is no longer at the cloud or like in federated or edge computing at a fusion center, but fully distributed at the device level. Synchrony requirements need not be assumed. Networked architectures may be viewed as a generalization of centralized and federated configurations, allowing us to recover federated algorithms from distributed \textcolor{black}{or decentralized} ones by employing a star-topology.

 Networked distributed processing architectures are more robust\textemdash if an edge or an agent fails, the entire network can continue to process data and deliver inference decisions. There is no need for costly communications with the cloud or a remote edge server. \textcolor{black}{Furthermore, while the exchange of processed iterates} \textcolor{black}{might still}  \textcolor{black}{leak some  private information, recent works have demonstrated that networked architectures can be designed to offer improved privacy due to their decentralized nature\nottoggle{short}{~\cite{Vlaski21:Graph-HomomorphicPerturbationForPrivateDecentralizedLearning, Wang22:DecentralizedStochasticOptimizationWithInherentPrivacy,pequito2014design,ramos2021distributed}}. Even more importantly, distributed networked architectures can be shown to match the performance of centralized solutions}.

This tutorial article surveys the recent history of networked signal and information processing including consensus and diffusion strategies for regression problems \cite{Rabbat04, Saber07, Lopes08, Schizas08, Kar09} developed in the 2000s, detection and parameter estimation over networks~\iftoggle{short}{\cite{Barbarossa07, Dimakis10, Kar12, Cattivelli10}}{\cite{Barbarossa07, Dimakis10, Kar12, Cattivelli10, Ribeiro06}} and their performance guarantees~\iftoggle{short}{\cite{Kar12, Cattivelli10, bajovic2012large, Matta16}}{\cite{Kar12, Cattivelli10, Ribeiro06, bajovic2012large, Matta16}}, distributed optimization~\iftoggle{short}{\cite{Nedic09, Boyd11, Ram10, Shi15, Ling15, jakoveticxaviermoura-TAC2014,jakoveticbajovicxaviermoura-Pr2020, DiLorenzo16, Sayed14proc, Chen15transient}}{\cite{Nedic09, Duchi12, Boyd11, Ram10, Mota13:D-AMMM, Ling14:DecentralizedDynamic, Shi15, Ling15, jakoveticxaviermoura-TAC2014,jakoveticbajovicxaviermoura-Pr2020, Chen13, DiLorenzo16, Sayed14proc, Chen15transient}}, learning, and adaptation~\cite{Chen15transient, Sayed14, Sayed14proc}. It provides a comprehensive coverage of topics and references. We will bridge the gap by unifying under a common umbrella more recent applications to distributed machine learning including multitask learning~\cite{Nassif20} and nonconvex optimization~\cite{Vlaski19nonconvexP2, Swenson21}, design variants under different operating scenarios such as asynchronous algorithms~\cite{Zhao15} and connections to alternative architectures such as federated learning\nottoggle{short}{~\cite{Kairouz21}}.

\vspace{-4mm}\section{Historical remarks}\vspace{-2mm}\label{sec:history}
There has been extensive work on distributed techniques for information and signal processing over networks. Many optimal inference problems adopt a quadratic optimization cost whose solution, under linear models and Gaussian noise, is a linear statistic of the data. With peer-to-peer communication among sensors, the question becomes how to compute the global average of the local statistics only through cooperation among the agents. Departing from centralized architectures, the solution built on the consensus strategy for distributed averaging, with no need for a fusion center to collect the dispersed data for processing. Consensus was initially proposed by DeGroot~\cite{degroot1974reaching} to  enable a group of agents to reach agreement by pooling their information and \textcolor{black}{to} converge~\cite{degroot1974reaching, Berger81} to an average estimate solely by interaction among neighbors. Many subsequent works were devoted to characterizing consensus' convergence behavior, the role of the graph topology, random selection of neighbors, and several other aspects. Some sample works include~\iftoggle{short}{\cite{Tsitsiklis86, Boyd-GossipInfTheory, Kar09}}{\cite{Tsitsiklis84, Tsitsiklis86, Xiao04, Boyd-GossipInfTheory, Kar09}}, while useful overviews appear in~\cite{Dimakis10, Sayed14} with many additional references.

Several works in the literature proposed extensions of the original consensus construction in order to more generally minimize aggregate cost functions, such as mean-square-error costs, or to solve distributed estimation problems of the least-squares or Kalman filtering type. These extensions involve constructions with gradient-descent type updates. Among these works we may mention~\iftoggle{short}{\cite{Tsitsiklis86, KM-JSTSP}}{\cite{Tsitsiklis86, olfati-saber:2005, Speranzon06, Schenato-Kalman, KM-JSTSP}}. While an early version of the consensus gradient-based algorithm for distributed estimation and optimization already appears in~\cite{Tsitsiklis86}, convergence results were limited to the asymptotic regime and there was no understanding of the performance of the algorithm, its actual convergence rate, and the influence of data, topology, quantization, noise, and asynchrony on behavior. These considerations are of great significance when designing practical, data-driven systems and they attracted the attention of the signal processing community after the turn of the century.  Moreover, some of the earlier investigations on consensus implementations involved separate time scales (fast communication and consensus iterations among agents, slow data collection), which can be a challenge for streaming or \textit{online} data.

Online consensus implementations where data is collected at every consensus step, appeared in the works by~\cite{Kar09, Braca08, Nedic09, karmoura-quantized, Kar12} and others. Using decaying step-sizes, these works  established the ability of the algorithms to converge. In particular, the work~\cite{Kar12} introduced the so-called consensus+innovations variant, which responds to streaming data and established  several performance measures in terms of convergence rate, and the effect of topology, quantization, and noisy conditions and other factors---see also~\cite{karmoura-quantized}. In parallel with these developments, online distributed algorithms of the diffusion type were introduced by~\iftoggle{short}{\cite{Sayed07, Lopes08, Cattivelli08}}{\cite{lopessayed-lincolnlab06, Sayed07, Lopes08, Cattivelli08}} to enable continuous adaptation and learning by networked agents under constant step-size conditions. The diffusion strategies modified the consensus update in order to incorporate symmetry, which was shown to enlarge the stability range of the network and to enhance its performance, even under decaying step-sizes---see~\cite{Towfic16} and the overviews~\cite{Sayed14, Sayed14proc}. The diffusion structure was used in several subsequent works for distributed optimization such as~\iftoggle{short}{\cite{Ram10, Srivastava11, Sayed22}}{\cite{Ram10, Srivastava11, Stankovic-parameter, Sayed22}} and other works.

In all these and related works on online distributed inference, the goal is for  every agent in the network to converge to an estimate of the unknown by relying exclusively on local interactions with its neighbors.  Important questions that arise in this context include: 1) \textit{convergence}: do the distributed inference algorithms converge and if so in what sense; 2) \textit{agreement}: do the agents reach a consensus on their inferences; 3) \textit{distributed versus} \textit{centralized}: how good is the distributed inference solution at each agent  when compared with the centralized inference obtained by a fusion center, in other words are the distributed inference sequences consistent, and asymptotically unbiased, efficient, and normal; and 4) \textit{rate of convergence}: what is the rate at which the distributed inference at each agent converges. These questions require very different approaches than the methods used in the ``consensus or averaging only'' solution from earlier works. Solutions that emerged of the consensus and diffusion type combine at each iteration 1) an \textit{aggregation step} that fuses the current inference statistic  at each agent with the states of their neighbors, with 2) a \textit{local update} driven by the new observation at the agent. This generic framework, of which there are several variations, is very different from the standard consensus where in each time step only local averaging of the neighbors' states occurs, and no observations are processed, and from other distributed inference algorithms with multiple time-scales, where between-measurement updates involve a large number of consensus steps (theoretically, an infinite number of steps). The classes of successful distributed inference algorithms that emerged add only to the identifiability condition of the centralized model that the network be connected on average. The results for these algorithms are also shown to hold under broad conditions like agents' communication channel intermittent failures, asynchronous and random communication protocols, and quantized communication (limited bandwidth), making their application realistic when 1) a large number of agents are involved (bound to fail at random times), 2) packet losses in wireless digital communications cause links to fail intermittently, 3) agents communicate asynchronously and 4) the agents may be resource constrained and have a limited bit budget for communication. Furthermore, these distributed inference algorithms make no distributional assumptions on the agents and link failures that can be spatially correlated. Readers may refer to the overviews~\cite{Dimakis10, Sayed14, Sayed14proc} and the many references therein.

There have of course been many other useful contributions to the theory and practice of networked information and signal processing, with many other variations and extensions. However, space limitations prevent us from elaborating further. Readers can refer to the overviews~\cite{Dimakis10, Sayed14}. Among such extensions, we may mention extensive works on distributed Kalman filtering by~\iftoggle{short}{\cite{Khan-Moura, Cattivelli10kalman, karmoura-kalman-2011}}{\cite{olfati-saber:2005, Khan-Moura, msechu2008decentralized, Cattivelli10kalman, karmoura-kalman-2011}} and others. Other parts of this manuscript refer to other classes of distributed algorithms, such as constructions of the primal and primal-dual type, and the corresponding references. The presentation actually presents a unified view of a large family of distributed implementations, including consensus and diffusion, for online inference by networked agents.

\textcolor{black}{\textbf{Notation.} All vectors are column vectors. We employ bold font to emphasize random variables, and regular font for their realizations as well as deterministic quantities. Upper case letters denote matrices, while Greek letters denote scalar variables. We will employ \( k \) to index nodes or agents in the network, and \( i \) to index time. In this way, \( \x_{k, i} \) will denote the data available to agent \( k \) at time \( i \), modeled as a random variable. When discussing supervised learning problems, \( \x_{k, i} \triangleq \mathrm{col}\{ \boldsymbol{h}_{k, i}, \boldsymbol{\gamma}_{k, i} \} \) will contain both the feature vector \( \boldsymbol{h}_{k, i} \) and the associated label \( \boldsymbol{\gamma}_{k, i} \).}

\vspace{-4mm}\section{Unified view}
\begin{figure}[htb]
  \centering
	\includegraphics[width=.7\columnwidth]{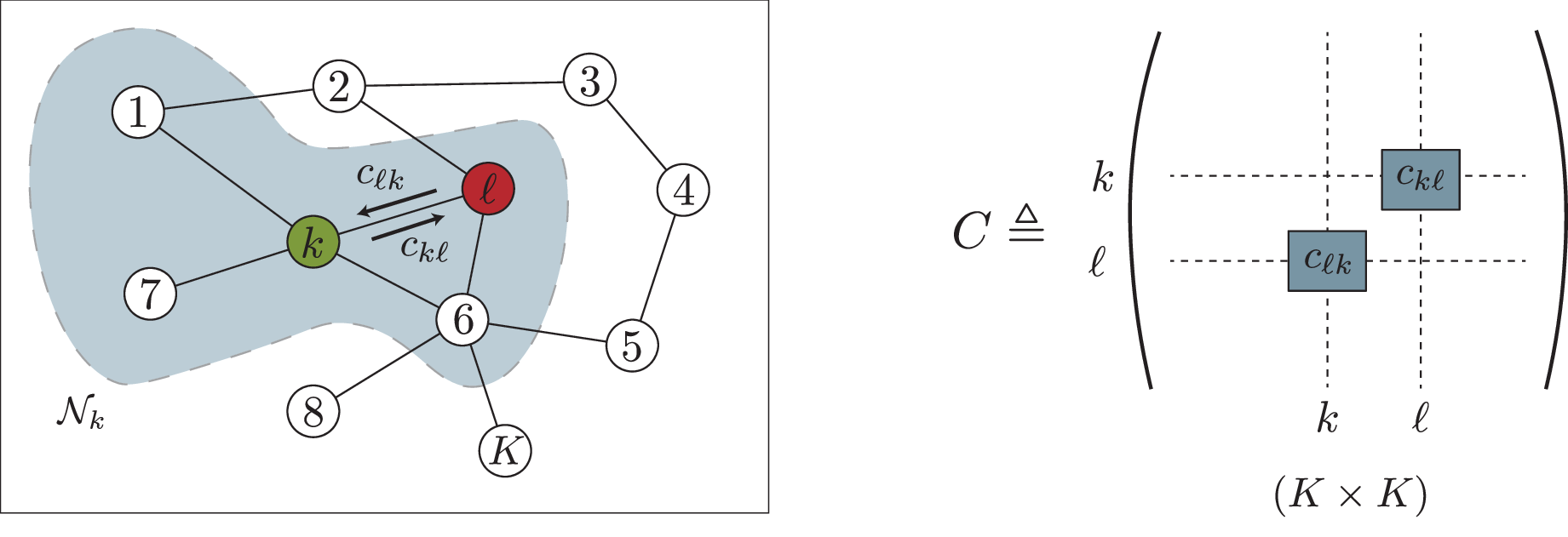}
\caption{\small Schematic of a general network and its adjacency matrix.}\label{fig:graph}
\end{figure}

\vspace{-2mm}\label{sec:unifiedview}
In the networked signal and information processing context, \( K \) agents are nodes of a connected network, whose graph is described by a weighted adjacency matrix \( C \in \mathds{R}^{K \times K} \), where \( c_{\ell k} \triangleq [C]_{\ell k} \) denotes the strength of the link from node \( \ell \) to node \( k \). We denote by $\mathcal{N}_k$ the neighbors of node~$k$, i.e., those other agents with which~$k$ communicates \textcolor{black}{directly} and cooperates. With undirected graphs, the graph is also described by its (weighted) Laplacian matrix\iftoggle{short}{, \( L = \mathrm{diag} \left\{ C \mathds{1} \right\} - C \).}{:
 \vspace{-1mm}\begin{align}\label{eq:laplacian_def}
    L = \mathrm{diag} \left\{ C \mathds{1} \right\} - C
 \end{align}}
 Here, \( \mathds{1} \) denotes the vector consisting of all ones of appropriate size. \textcolor{black}{We illustrate an example of a graph and its adjacency matrix in Fig.~\ref{fig:graph}.} We further associate with each node a local model \( w_k \in \mathds{R}^M \), which can correspond to \textcolor{black}{unknown} parameters \textcolor{black}{describing} a \textcolor{black}{random field, or parameterizing a} channel \textcolor{black}{or} filter, \textcolor{black}{or representing a} hyperplane or a neural network. For convenience, we define \emph{network-level quantities}, which we denote through calligraphic letters; they aggregate quantities from across the network. In this manner, we can write compactly\iftoggle{short}{ \( \cw \triangleq \mathrm{col} \left\{ w_k \right\} \).}{:
 \vspace{-1mm}\begin{align}
    \cw \triangleq \mathrm{col} \left\{ w_k \right\}
 \end{align}}
 This notation allows us to highlight a useful relation between the adjacency matrix \( C \), and the Laplacian matrix \( L \), namely that \textcolor{black}{for undirected graphs}: \vspace*{-1mm}\begin{align}\label{eq:variation_measure}
    \sum_{k=1}^K \sum_{\ell = 1}^K c_{\ell k} \|w_k - w_{\ell}\|^2 = \cw^{\mathsf{T}} \mathcal{L} \cw
 \end{align}
where we defined \( \mathcal{L} \triangleq L \otimes I_M \). Relation~\eqref{eq:variation_measure} \textcolor{black}{captures through the variational operator~$\mathcal{L}$} the weighted variation of the local models \( w_k \) across the network, and is fundamental when deriving algorithms for distributed processing over networks, as we illustrate further below.
\nottoggle{short}{\textcolor{black}{\begin{example}[\textbf{A numerical example}]\label{ex:line_graph} Let us consider a graph consisting of three nodes, arranged in \textcolor{black}{an undirected} line graph with symmetric weight matrix:
\begin{align}
  C = \begin{pmatrix} 1 & 1 & 0 \\ 1 & 1 & 1 \\ 0 & 1 & 1 \end{pmatrix} \Longleftrightarrow L \stackrel{\eqref{eq:laplacian_def}}{=} \begin{pmatrix} 2 & 0 & 0 \\ 0 & 3 & 0 \\ 0 & 0 & 2 \end{pmatrix} - \begin{pmatrix} 1 & 1 & 0 \\ 1 & 1 & 1 \\ 0 & 1 & 1 \end{pmatrix} = \begin{pmatrix} 1 & -1 & 0 \\ -1 & 2 & -1 \\ 0 & -1 & 1 \end{pmatrix}
\end{align}
To simplify calculations, consider scalar weight vectors \( w_k \in \mathds{R} \) and two instances \textcolor{black}{of~$\cw$}:
\begin{align}
  \{ w_1 = 1; w_2 = 2; w_3 = 3 \} \Longrightarrow \cw = \begin{pmatrix} 1 \\ 2 \\ 3 \end{pmatrix} \ \mathrm{or}\ \{ w_1' = 2; w_2' = 1; w_3' = 3 \} \Longrightarrow \cw' = \begin{pmatrix} 2 \\ 1 \\ 3 \end{pmatrix}
\end{align}
\begin{figure}[H]
  \centering
	\includegraphics[width=.4\columnwidth]{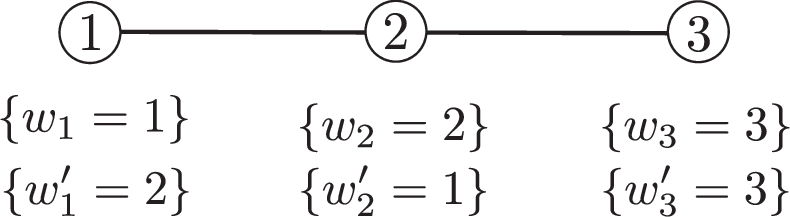}
\caption{\small Schematic of the line graph from Example~\ref{ex:line_graph} with associated weights \( \cw \) and \( \cw' \).}\label{fig:line_graph}
\end{figure}
\noindent We illustrate the line graph and associated weight vectors in Fig.~\ref{fig:line_graph}. Note that both \( \cw \) and \( \cw' \) have the same energy, i.e., \( \|\cw\|^2 = \|\cw'\|^2 = 14\). Nevertheless, if we compute the variation measure~\eqref{eq:variation_measure}, we find:
\begin{align}
  {\cw}^{\mathsf{T}} \mathcal{L} \cw = 2 \ \mathrm{and}\ {\cw'}^{\mathsf{T}} \mathcal{L} \cw' = 5
\end{align}
and hence \( {\cw}^{\mathsf{T}} \mathcal{L} \cw < {\cw'}^{\mathsf{T}} \mathcal{L} \cw' \). This result quantifies the fact that the weight vectors \( \{ w_1 = 1; w_2 = 2; w_3 = 3 \} \) vary more smoothly than \( \{ w_1' = 2; w_2' = 1; w_3' = 3 \} \) over the line graph described by the weight matrix \( C \) or Laplacian matrix \( L \), where node \( 2 \) is central.
\end{example}
\begin{example}[A visual example] To illustrate the graphical variation measure~\eqref{eq:variation_measure} on a larger graph, we place \( K = 1000 \) nodes uniformly at \textcolor{black}{random} in a two-dimensional plane, and construct a graph \( C \) and Laplacian \( L\) geometrically by \textcolor{black}{connecting nodes whose} 
 Euclidean distance \textcolor{black}{is smaller than a threhshold}. The weight \( c_{\ell k} \) are chosen proportional to the proximity of nodes \( \ell \) and \( k \). We then generate \( \cw \) by sampling from a Gaussian Markov random field with distribution
  \begin{align}\label{eq:gmrf}
    f(\cw) = {(2\pi\eta)}^{-M(K-1)/2} \left( |\mathcal{L}|^{*} \right)^{1/2} e^{-\frac{\eta}{2} \cw^{\mathsf{T}} \mathcal{L} \cw}
  \end{align}
where \( |\cdot|^{*} \) denotes the pseudo-determinant of a matrix, i.e., the product of non-zero eigenvalues. Here \( \eta \) represents a temperature parameter and smaller \( \eta \) is more likely to generate weight vectors \( w_k \) with large graphical variability. We show the resulting distributions in Fig.~\ref{fig:gmrf} for varying \( \eta \).
\begin{figure}[H]
  \begin{minipage}{.32\textwidth}
    \centering    \includegraphics[width=\linewidth]{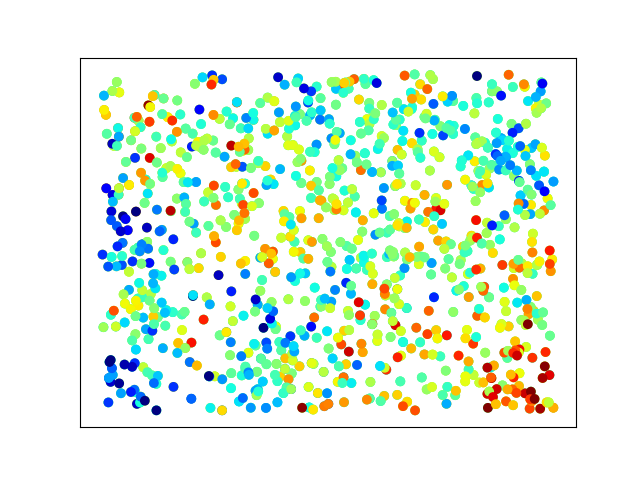}
    \begin{center}\( \eta = 0.001 \), large \( \cw^{\mathsf{T}} \mathcal{L} \cw \)\end{center}
  \end{minipage}
  \begin{minipage}{.32\textwidth}
    \centering    \includegraphics[width=\linewidth]{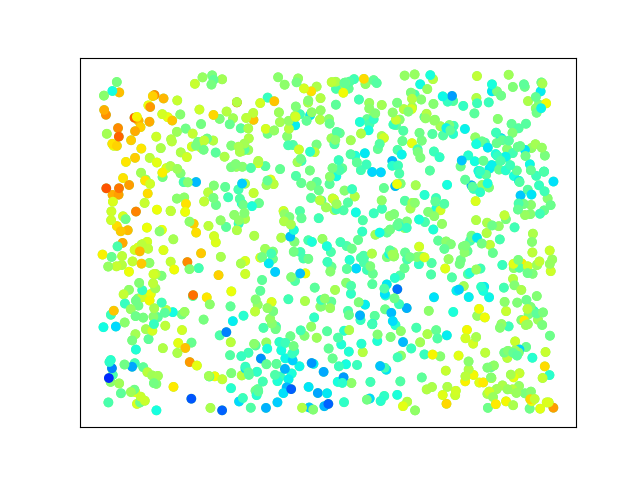}
    \begin{center}\( \eta = 0.005 \), moderate \( \cw^{\mathsf{T}} \mathcal{L} \cw \)\end{center}
  \end{minipage}
  \begin{minipage}{.32\textwidth}
    \centering    \includegraphics[width=\linewidth]{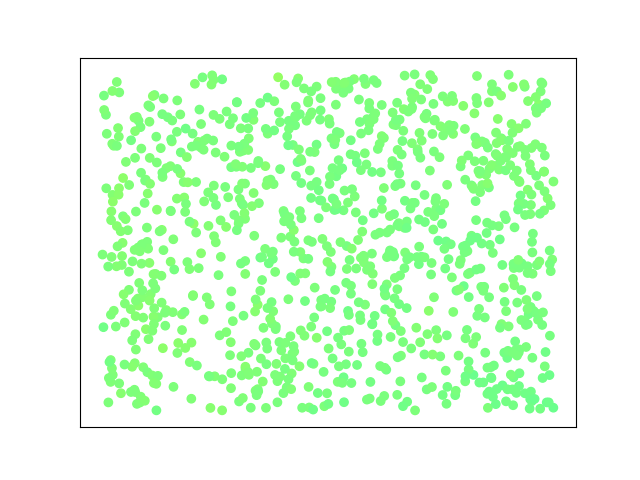}
    \begin{center}\( \eta = 1 \), small \( \cw^{\mathsf{T}} \mathcal{L} \cw \)\end{center}
  \end{minipage}
  \caption{\small \textcolor{black}{Gauss Markov random field samples.}\label{fig:gmrf}}
\end{figure}
\end{example}}}

\subsection{Unification through Stochastic Optimization}\vspace{-2mm}\label{sec:derivations}
\textcolor{black}{Suppose we would like each agent in the network to estimate the \textcolor{black}{unknown} parameter \( w_k^o \) \textcolor{black}{used to generate local observations through the linear model} 
\vspace*{-1.5mm} \begin{align}\label{eqn:linearobsmodel}
\boldsymbol{\gamma}_{k, i} = \boldsymbol{h}_{k, i}^{\mathsf{T}} w_k^o + \boldsymbol{v}_{k, i} \end{align}
\textcolor{black}{In a parameter estimation framework,} \( \boldsymbol{h}_{k, i} \) denotes the \textcolor{black}{local known observation model,}  \( \boldsymbol{v}_{k, i} \) denotes noise, and \( \boldsymbol{\gamma}_{k, i} \) \textcolor{black}{are the observations}. \textcolor{black}{In a machine learning interpretation, during training, we learn the weights or model~$w_k^o$ in~\eqref{eqn:linearobsmodel} from known pairs of input data \( \boldsymbol{h}_{k, i} \) and target values  \( \boldsymbol{\gamma}_{k, i} \). Common other terminology refers to \( \boldsymbol{h}_{k, i} \) as regressor,  feature vector, independent variables, or inputs}. We may then formulate local \textcolor{black}{estimation or} learning problems based on the mean squared error (MSE) risk, \textcolor{black}{\( J_k(w_k) \triangleq \mathds{E} \| \boldsymbol{\gamma}_{k, i} - \boldsymbol{h}_{k, i}^{\mathsf{T}} w_k\|^2, \)} and pursue:
\vspace*{-1.5mm}
\begin{align}\label{eq:lin_obs_cost_model}
  \textcolor{black}{w_{k}^o \triangleq \arg\min_{w_k \in \mathds{R}^M} J_k(w_k) \hspace*{.5cm}\textrm{or equivalently}\hspace*{.5cm}}\min_{\cw} \sum_{k=1}^K J_k(w_k)
\end{align}
If, however, we are provided with prior information that the parameters \( w_k^o \) vary smoothly as defined by the \textcolor{black}{variational relation}~\eqref{eq:variation_measure} over a graph with Laplacian \( L \), we may instead pursue:
\begin{align}\label{eq:mse_example}
  \min_{\cw} \left\{ \sum_{k=1}^K J_k(w_k) + \frac{\eta}{2} \cw^{\mathsf{T}} \mathcal{L} \cw \right\}
\end{align}
The regularization term \( \frac{\eta}{2} \cw^{\mathsf{T}} \mathcal{L} \cw \) couples the independent objectives \( J_k(w_k) \) and encourages parameterizations \( w_k \) that vary smoothly over the graph. It can be verified that the coupled optimization problem~\eqref{eq:mse_example} corresponds to a maximum aposteriori estimate of the models \( w_k^o \) in the linear model \( \eqref{eqn:linearobsmodel}\) \iftoggle{short}{under a Gaussian Markov random field prior. Motivated by applications in wireless sensor networks, least-squares problems of this form were the focus of many of the early works on distributed processing~\cite{Lopes08, Schizas08,Kar09,kar2011convergence,Cattivelli10}}{under the GMRF prior~\eqref{eq:gmrf}}.}

\textcolor{black}{More generally, with} each agent, we associate a local objective function \( J_k(w_k) = \mathds{E} Q(w_k; \boldsymbol{x}_{k, i}) \), where \( \boldsymbol{x}_{k, i} \) refers to the data that is available to agent \( k \) and \( Q(w_k; \boldsymbol{x}_{k, i}) \) is a loss function. \textcolor{black}{Setting \( Q(w_k; \boldsymbol{x}_{k, i}) \triangleq \| \boldsymbol{\gamma}_{k, i} - \boldsymbol{h}_{k, i}^{\mathsf{T}} w_k\|^2 \) recovers the MSE loss leading to~\eqref{eq:mse_example}.} We consider the general class of coupled optimization problems:
\vspace{-1mm}\begin{align}\label{eq:generic_couple}
  \min_{\cw \in \mathds{R}^{K M}} \left\{ \sum_{k=1}^K J_k(w_k) + \eta R\left(\cw \right) \right\}, \ \textrm{subject to } \cw \in \Omega
\end{align}
The coupling regularizer \( R\left( \cw \right) \) and constraint \( \cw \in \Omega \) encode relationships between local objectives and encourages local cooperation. \textcolor{black}{Letting \( R\left( \cw \right) = \frac{1}{2} \cw^{\mathsf{T}} \mathcal{L} \cw \) and \( \Omega = \mathds{R}^{KM} \) recovers the smoothed aggregate learning problem~\eqref{eq:mse_example}.} While decentralized algorithms for learning and optimization can be developed for general asymmetric adjacency matrices \( C \neq C^{\mathsf{T}} \) \cite{Sayed14, xi2016distributed}, for the sake of simplicity, we will focus on symmetric adjacency matrices in this section. We will comment on the implications of employing asymmetric combination policies in Sec.~\ref{sec:asymmetric}.

\nottoggle{short}{\vspace{-4mm}\subsubsection{Model and Cost Formulation}\vspace{-2mm}
For illustration purposes, consider a networked denoising problem, where each agent \( k \) corresponds to a sensor observing some unknown parameter \( w_k^o \) under noise, i.e., \( \boldsymbol{\gamma}_{k, i} = w_k^o + \boldsymbol{v}_{k, i} \). We may then formulate the local mean square denoising problem \(
  J_k(w_k) \triangleq \mathds{E} \|\boldsymbol{\gamma}_{k, i} - w_k \|^2 \).
Instead of observing \( w_k^o \) directly under noise, we may observe a linear and underdetermined transformation of \( w_k^o \) under the model \( \boldsymbol{\gamma}_{k, i} = \boldsymbol{h}_{k, i}^{\mathsf{T}}w_k^o + \boldsymbol{v}_{k, i} \), and formulate instead the mean square estimation problem \(
  J_k(w_k) \triangleq \mathds{E} \|\boldsymbol{\gamma}_{k, i} - \boldsymbol{h}_{k, i}^{\mathsf{T}}w_k\|^2 \). Motivated by applications in wireless sensor networks, least-squares problems of this form were the focus of many of the early works on distributed processing~\cite{Lopes08, Schizas08,Kar09,kar2011convergence,Cattivelli10}.

  We may also consider other models and costs, such as logistic regression. In this setting, appropriate in the case of discrete labels \( \boldsymbol{\gamma}_{k, i} \in \{ \pm 1 \} \), the risk function is chosen as \(
  J_k(w_k) \triangleq \mathds{E} \ln\left( 1 + e^{- \boldsymbol{\gamma}_{k, i} \boldsymbol{h}_{k, i}^{\mathsf{T}} w_k} \right) + \frac{\rho}{2}\|w_k\|^2\). The empirical success of deep learning based approaches~\cite{Goodfellow16, Sayed22} has driven increasing interest in the deployment of nested models, which allow for the learning of non-linear relationships, both for estimation and classification. The general framework~\eqref{eq:generic_couple} is amenable to such settings by letting \(
  J_k(w_k) \triangleq \mathds{E}\left \| \boldsymbol{\gamma}_{k, i} - \sigma\left( {W}_{k, L} \cdot \sigma \left( {W}_{k, L-1} \cdot\cdot\cdot \sigma\left( {W}_{k, 1} \boldsymbol{h}_{k, i} \right) \right) \right) \right\|^2 \),
where the parameter vector \( w_k \) now corresponds to the aggregate of the layer-wise parameters \( \{ W_{k, \ell} \}_{\ell =1}^L \) and \( \sigma(\cdot) \) denotes some non-linear activation function. Variations based on the cross-entropy loss, which is more appropriate in the context of classification problems, are also possible, as are formulations based on more elaborate neural network architectures, such as convolutional or recurrent neural networks~\cite{Goodfellow16, Sayed22}.}

\vspace{-4mm}\subsubsection{Stochastic Gradient Approximations}\vspace{-2mm}\label{sec:stochastic_approximations}
A common theme in many networked data processing applications is the limited access to the cost function \( J_k(\cdot) \) and its gradient \( \nabla J_k(\cdot) \), due to the fact that the cost \( J_k(\cdot) \) is defined as the expected value of the loss \( Q_k(\cdot; \x_{k, i}) \) and \( \x_{k, i} \) follows some \emph{unknown} distribution. As a result, gradient descent algorithms that rely on the use of the true gradient \( \nabla J_k(\cdot) \), are replaced by stochastic gradient algorithms, which employ an approximated gradient denoted by \( \widehat{\nabla J}_k(\cdot)\). The most common construction for a stochastic gradient approximation is to employ\iftoggle{short}{ \( \widehat{\nabla J}_k(\cdot) \triangleq \nabla Q(\cdot; \x_{k, i}) \),}{:
\vspace{-1mm}\begin{align}\label{eq:construction_sgd}
  \widehat{\nabla J}_k(\cdot) \triangleq \nabla Q(\cdot; \x_{k, i})
\end{align}}
where \( \x_{k, i} \) denotes a single sample of the variable \( \x_{k} \) obtained at time \( i \). However, other constructions are possible depending on the setting. For example, we may envision a scenario where agent \( k \) is provided with several independent samples \( \{\x_{k, i, b}\}_{b=1}^{B_k} \) at time \( i \), allowing for the mini-batch construction\iftoggle{short}{ \( \widehat{\nabla J}_k(\cdot) \triangleq \frac{1}{B_k}\sum_{b=1}^{B_k} \nabla Q(\cdot; \x_{k, i, b}) \).}{:
\vspace{-1mm}\begin{align}\label{eq:construction_minibatch}
  \widehat{\nabla J}_k(\cdot) \triangleq \frac{1}{B_k}\sum_{b=1}^{B_k} \nabla Q(\cdot; \x_{k, i, b})
\end{align}}
Alternatively, one may be faced with a situation where agents may be able to provide a gradient approximation only with some probability \( \pi_{k} \), either due to lack of data, slow or delayed updates, or computational failure. Such asynchronous behavior can be modeled via~\cite{Zhao15}:
\vspace{-1mm}\begin{align}\label{eq:construction_async}
  \widehat{\nabla J}_k(\cdot) = \begin{cases} \frac{1}{\pi_k}  \nabla Q(\cdot; \x_{k, i}), &\textrm{with prob. \( \pi_k \),}\\ 0, &\textrm{with prob. \( 1 - \pi_k \).} \end{cases}
\end{align}
As a final example of commonly used constructions for stochastic gradient approximations, we note perturbed stochastic gradients of the form\iftoggle{short}{ \( \widehat{\nabla J}_k(\cdot) = \nabla Q(\cdot; \x_{k, i}) + \textcolor{black}{\boldsymbol{r}_{k, i}} \),}{:
\vspace{-1mm}\begin{align}\label{eq:construction_noisy}
  \widehat{\nabla J}_k(\cdot) = \nabla Q(\cdot; \x_{k, i}) + \textcolor{black}{\boldsymbol{r}_{k, i}}
\end{align}}
where \( \textcolor{black}{\boldsymbol{r}_{k, i}} \) denotes some additional zero-mean noise. Examples of settings where additional noise is added to gradient approximations are plentiful, and include noise added due to quantization, noise used to ensure differential privacy\nottoggle{short}{~\cite{Dwork14}}, or noise used to escape from saddle-points in nonconvex environments~\iftoggle{short}{\cite{Vlaski19nonconvexP2}}{\cite{Ge15, Vlaski19nonconvexP2}}. As we will see in the learning guarantees that we survey further ahead, the performance of the algorithms based on stochastic gradient approximations will in some way depend on the quality of \( \widehat{\nabla J}_k(\cdot) \). Most commonly this is quantified through bounds on its variance.
\begin{condition}{\textbf{(Variance of the Gradient Approximation).}}
  The gradient approximation \( \widehat{\nabla J}_k(\w_{k, i-1})\) is required to be unbiased with bounded variance as follows:
  \vspace{-1mm}\begin{align}
    \E \left\{ \widehat{\nabla J}_k(\w_{k, i-1}) | \w_{k, i-1} \right\} =&\: {\nabla J}_k(\w_{k, i-1}) \label{eq:zero_mean_cond}\\
    \E \left\{ {\left\|\widehat{\nabla J}_k(\w_{k, i-1}) - {\nabla J}_k(\w_{k, i-1}) \right\|}^2 | \w_{k, i-1} \right\} \le&\: \beta_k^2 \|w_k^o - \w_{k, i-1}\|^2 + \sigma_k^2 \label{eq:variance_bound_error}
  \end{align}
  Here, \( \beta_k^2, \sigma_k^2 \) denote non-negative constants, and \( w_k^o \) denotes an arbitrary reference point, \textcolor{black}{most commonly the minimizing argument from~\eqref{eq:lin_obs_cost_model}}.\qed
\end{condition}
\noindent\textcolor{black}{As already shown in~\cite{Sayed22}, the zero-mean condition~\eqref{eq:zero_mean_cond} can be verified to hold} for many popular constructions, including\iftoggle{short}{ the constructions listed above}{~\eqref{eq:construction_sgd}--\eqref{eq:construction_noisy}}. \textcolor{black}{In~\eqref{eq:zero_mean_cond} and~\eqref{eq:variance_bound_error}, we condition on the current iterate \( \w_{k, i-1} \) and take expectation with respect to the remaining variability in generating the gradient approximation \( \widehat{\nabla J}_k(\w_{k, i-1}) \), which is the data available to agent \( k \) at time \( i \). For example, in the case of ordinary stochastic gradient descent\iftoggle{short}{ \( \widehat{\nabla J}_k(\cdot) \triangleq \nabla Q(\cdot; \x_{k, i}) \)}{~\eqref{eq:construction_sgd}}, this corresponds to \( \x_{k, i} \), which is generally assumed to be independent of \( \w_{k, i-1} \).} Variance bounds of the form~\eqref{eq:variance_bound_error}, on the other hand, need to be verified for specific choices of loss functions \( Q(\cdot; \boldsymbol{x}_k) \), distributions of the data \( \x_{k, i} \), and gradient approximations \( \widehat{\nabla J}_k(\cdot) \). Nevertheless, the key take-away is that conditions of this form hold for most processing and learning problems of interest. The resulting constants \( \beta_k^2, \sigma_k^2 \) quantify the quality of the utilized gradient approximation. We list in Table~\ref{tab:gradient_noise} the relevant quantities for the mean-square error and logistic loss as examples.
 It is also useful to note that, given the constants \( \beta_k^2, \sigma_k^2 \) for an ordinary gradient approximation\nottoggle{short}{~\eqref{eq:construction_sgd}}, such as those listed in Table~\ref{tab:gradient_noise}, one can immediately recover those of the variants\iftoggle{short}{ listed above}{~\eqref{eq:construction_minibatch}--\eqref{eq:construction_noisy}}. This is illustrated in Table~\ref{tab:variations}.
\begin{table}[H]
  \centering
  \begin{tabular}{c|c|c|c}
    Loss &  Gradient Approximation & \( \beta_k^2 \) (relative) & \( \sigma_k^2 \) (absolute) \\
    \hline
    \( \frac{1}{2} \|\boldsymbol{\gamma}_{k, i} - \boldsymbol{h}_{k, i}^{\mathsf{T}} w\|^2 \) & \( \boldsymbol{h}_{k, i} \left( \boldsymbol{\gamma}_{k, i} - \boldsymbol{h}_{k, i}^{\mathsf{T}} w \right) \) & \( \E \|R_h - \boldsymbol{h}_{k}\boldsymbol{h}_{k}^{\mathsf{T}}\|^2 \) & \(\sigma_v^2 \mathrm{Tr}(R_h) \) \\
    \( \ln\left( 1 + e^{- \boldsymbol{\gamma}_{k, i} \boldsymbol{h}_{k, i}^{\mathsf{T}} w} \right) + \frac{\rho}{2} \|w\|^2 \) & \( \boldsymbol{\gamma}_{k, i} \boldsymbol{h}_{k, i} \left( \frac{1}{1 + e^{\boldsymbol{\gamma}_{k, i} \boldsymbol{h}_{k, i}^{\mathsf{T}} w}} \right) + \rho w \) & \( 0 \) & \( \mathrm{Tr}(R_h) \)
  \end{tabular}\caption{\small Constants in gradient variance bounds for popular loss functions for supervised learning problems with \( \boldsymbol{x}_k \triangleq \mathrm{col}\{ \boldsymbol{h}_k, \boldsymbol{\gamma}_{k} \} \)~\cite{Sayed14}. \textcolor{black}{The quantities \( \sigma_v^2 \) and \( R_h \) denote the data statistics \( \mathds{E} \boldsymbol{v}_{k, i}^2 \) and \( \mathds{E} \boldsymbol{h}_{k, i}\boldsymbol{h}_{k, i}^{\mathsf{T}} \) respectively.}}\label{tab:gradient_noise} 
\end{table}
\begin{table}[H]
  \centering
  \begin{tabular}{c|c|c|c}
    \( \widehat{\nabla J_k}(\w_{k, i-1}) \) & unbiased? & \( \beta_k^2\) (relative) & \( \sigma_k^2 \) (absolute) \\
    \hline
    \( \nabla Q(\w_{k, i-1}, \boldsymbol{x}_{k, i}) \) & yes & \( \beta_{k, \textrm{ord}}^2 \) & \( \sigma_{k, \textrm{ord}}^2 \) \\
    \( \frac{1}{B} \sum_{b=1}^B \nabla Q(\w_{k, i-1}, \boldsymbol{x}_{k, i, b}) \) & yes & \( \frac{\beta_{k, \textrm{ord}}^2}{B} \) & \( \frac{\sigma_{k, \textrm{ord}}^2}{B} \) \\
    Asynchronous as in~\eqref{eq:construction_async} & yes & \( \frac{\beta_{k, \textrm{ord}}^2}{\pi_k} + \frac{1-\pi_k}{\pi_k} \delta_k^2 \) & \( \frac{\sigma_{k, \textrm{ord}}^2}{\pi_k} \) \\
    \( \widehat{\nabla J}_k(\cdot) = \nabla Q(\cdot; \x_{k, i}) + \textcolor{black}{\boldsymbol{r}_{k, i}} \) & yes & \( \beta_{k, \textrm{ord}}^2 \) & \( \sigma_{k, \textrm{ord}}^2 + \textcolor{black}{\sigma_{r, k}^2} \) \\
  \end{tabular}\caption{\small\textcolor{black}{The quantities \( \beta_{k, \mathrm{ord}}^2, \sigma_{k, \mathrm{ord}}^2\) correspond to the gradient noise constants of the ``ordinary'' gradient approximation \( \widehat{\nabla J_k}(\w_{k, i-1}) = \nabla Q(\w_{k, i-1}, \boldsymbol{x}_{k, i})\), and can be read from Table~\ref{tab:gradient_noise}. Constants of the variants are provided in terms of the baseline quantities \(\beta_{k, \mathrm{ord}}^2, \sigma_{k, \mathrm{ord}}^2\).} The parameter \( \delta_k \) corresponds to the Lipschitz constant of the gradient \( \nabla J_k(\cdot)\).}\label{tab:variations}
\end{table}
\textcolor{black}{We finally note that the current exposition mainly focuses on methods that assume first-order (i.e., gradient type) information is available or accessible in the construction of the distributed algorithms. Due to intractability of gradient computation in certain applications (for instance, in scenarios where the cost model is not directly available but perhaps may be computed at desired query points via noisy simulations), one can resort to zeroth-order approaches. In this case, noisy and biased gradient estimates obtained from measuring function values using various difference approximations are used in the algorithm design in lieu of exact or unbiased gradients as assumed in the first-order setting --- see~\cite{Sayed22, sahu2020decentralized} and the references therein for more details.}

\vspace{-4mm}\subsubsection{Task Relationships}\vspace{-2mm}\label{sec:penalty_based}
As a separate consideration from the choice of the risk functions \( J_k(w_k) \), one may consider various frameworks for the relation between individual models \( w_k \), also referred to as tasks. In the absence of coupling regularization or constraints, i.e., in the case \textcolor{black}{the regularizer}  \( R\left( \{ w_k \}_{k=1}^K \right) = 0 \) and \( \Omega = \mathds{R}^{KM} \), optimization over the aggregate cost \( \sum_{k=1}^K J_k(w_k) \) decouples into independent problems \( J_k(w_k) \) over local models \( w_k \). These can be pursued in a non-cooperative manner.

Perhaps the most commonly studied framework for distributed optimization is that of consensus optimization, where individual models are required to be common, i.e., \( w_k = w \), giving rise to:
\vspace{-1mm}\begin{align}\label{eq:consensus_optimization_problem}
  \min_{ w }\sum_{k=1}^K J_k(w)
\end{align}
Networked algorithms for~\eqref{eq:consensus_optimization_problem} can be developed from~\eqref{eq:generic_couple} in several ways, giving rise to different families of algorithms for distributed optimization~\cite{Sayed22}, as we proceed to show.

\noindent \textbf{Penalty-based approaches}: We may encourage consensus by penalizing pairwise differences between connected agents, i.e., \( R\left( \{ w_k \} \right) = \frac{1}{2}\sum_{k=1}^K \sum_{\ell \in \mathcal{N}_k} c_{\ell k} \|w_k - w_{\ell}\|^2 \), resulting in:
\vspace{-1mm}\begin{align}\label{eq:penalized_general}
  \min_{\{ w_k \}_{k=1}^K} \left\{ \sum_{k=1}^K J_k(w_k) + \frac{\eta}{2}\sum_{k=1}^K \sum_{\ell \in \mathcal{N}_k} c_{\ell k} \|w_k - w_{\ell}\|^2 \right\} \Longleftrightarrow \min_{\cw} \left\{ \mathcal{J}(\cw) + \frac{\eta}{2} \cw^{\mathsf{T}} \mathcal{L} \cw \right\}
\end{align}
where in addition to making use of~\eqref{eq:generic_couple}, we defined \( \mathcal{J}(\cw) \triangleq \sum_{k=1}^K J_k(w_k) \). It can be verified that, as long as the graph described by \( C \) is connected, \( \frac{1}{2}\sum_{k=1}^K \sum_{\ell \in \mathcal{N}_k} c_{\ell k} \|w_k - w_{\ell}\|^2 = 0 \), if and only if, \( w_k = w \) for all \( k \), and hence~\eqref{eq:penalized_general} is equivalent to~\eqref{eq:consensus_optimization_problem} in the limit as \( \eta \to \infty \). At the same time, this fact implies that for finite \( \eta \), problems~\eqref{eq:penalized_general} and~\eqref{eq:consensus_optimization_problem} will in general have distinct solutions. It is for this reason that penalty-based methods generally operate with large choices of the penalty parameter \( \eta \), exhibit\textcolor{black}{ing} some small bias relative to the exact consensus problem~\eqref{eq:consensus_optimization_problem}, unless \( \eta \to \infty \). \nottoggle{short}{

}Applying stochastic gradient descent to~\eqref{eq:penalized_general} results in:
\vspace{-1mm}\begin{align}
  \bcw_i = \left( I - \mu \eta \mathcal{L} \right) \bcw_{i-1} - \mu \widehat{\nabla \mathcal{J}}(\bcw_{i-1})
\end{align}
\textcolor{black}{If we set 
\( A \triangleq  I - \mu \eta \mathcal{L} \)} and return to node-level quantities, we recover the recursion:
\vspace{-1mm}\begin{align}\label{eq:consensus+innovation}
  \w_{k, i} = \sum_{\ell \in \mathcal{N}_k} a_{\ell k} \w_{\ell, i-1} - \mu \widehat{\nabla J}_k(\w_{k, i-1})
\end{align}
which corresponds to the decentralized (stochastic) gradient descent algorithm~\cite{Tsitsiklis86, Nedic09} of the ``consensus + innovation'' type~\cite{Kar12}. If we instead, following~\cite{Yuan18}, appeal to an incremental gradient descent argument, where we first take a step relative to the cost \( \mathcal{J}(\cw) \), and subsequently descend along the penalty \( \frac{\eta}{2} \cw^{\mathsf{T}} \mathcal{L} \cw \), we obtain the adapt-then-combine (ATC) diffusion algorithm~\iftoggle{short}{\cite{Lopes08, Chen15transient}}{\cite{Lopes08, Chen13, Chen15transient}}:
\vspace{-1mm}\begin{align}
  \boldsymbol{\psi}_{k, i} =&\: \w_{k, i} - \mu \widehat{\nabla J}_k(\w_{k, i-1})\label{eq:adapt} \\
  \w_{k, i} =&\: \sum_{\ell \in \mathcal{N}_k} {a}_{\ell k} \boldsymbol{\psi}_{\ell, i}\label{eq:combine}
\end{align}
Reversing the order in the argument yields instead the combine-then-adapt (CTA) variation of diffusion~\cite{Lopes08, Chen15transient}. \nottoggle{short}{We note that we denote by \( A \) the adjacency matrix guiding the flow of information such as in~\eqref{eq:consensus+innovation} and~\eqref{eq:combine}, while \( C \) parametrizes the regularization in~\eqref{eq:penalized_general}. In general, \( A \) will follow from \( C \) and the corresponding Laplacian matrix \( L \), though specific relations depending on the algorithm at hand.}

\nottoggle{short}{\begin{example}[\textbf{Graph Signal Denoising}] It is instructive to specialize~\eqref{eq:penalized_general} to the denoising problem in order to highlight relations with the field of graph signal processing~\cite{Ortega18}. In this case we consider the formulation:
\vspace{-1mm}\begin{align}
  \min_{\{ w_k \}_{k=1}^K} \frac{1}{2}\sum_{k=1}^K \mathds{E} \|\boldsymbol{\gamma}_{k, i} - w_k\|^2 + \frac{\eta}{2} \cw^{\mathsf{T}} \mathcal{L} \cw
\end{align}
which can be viewed as a regularized estimate of \( \{ w_k^o \}_{k=1}^K \) given measurements \( \boldsymbol{\gamma}_{k, i} = w_k^o + \boldsymbol{v}_{k, i} \) and a Gaussian Markov random field prior over \( \mathrm{col}\{ \boldsymbol{\gamma}_{k, i} \}_{k=1}^K \). Its closed form solution is given by \( \left( I + \eta \mathcal{L} \right)^{-1} \mathrm{col}\{ \E \boldsymbol{\gamma}_{k, i} \}\), which corresponds to a lowpass filtered version of the expected observations \( \E \boldsymbol{\gamma}_{k, i} \) in the graph domain induced by the Laplacian \( \mathcal{L} \). The regularization constant \( \eta \) controls the steepness of the lowpass filter, and we recover in the limit as \( \eta \to \infty \) the DC-signal \( \frac{1}{K} \sum_{k=1}^K \E \boldsymbol{\gamma}_{k, i} \) at all nodes. Evaluating \( \left( I + \eta \mathcal{L} \right)^{-1} \mathrm{col}\{ \E \boldsymbol{\gamma}_{k, i} \} \) over a network is not feasible since \( (a) \) \( \left( I + \eta \mathcal{L} \right)^{-1} \) is generally a dense matrix, requiring communication with a fusion center, and \( (b) \) it requires knowledge of the means \( \E \boldsymbol{\gamma}_{k, i} \). If we instead appeal to a construction of the form~\eqref{eq:consensus+innovation} but leave \( \mu \) and \( \eta \) uncoupled, we recover:
\vspace{-1mm}\begin{align}
  \w_{k, i} = \sum_{\ell \in \mathcal{N}_k} (1 - \mu - \mu \eta \mathcal{L}_{[k \ell]})\w_{\ell, i-1} + \mu \boldsymbol{\gamma}_{k, i}
\end{align}
which corresponds to an IIR graph filter that can be implemented in a distributed fashion using realizations of \( \boldsymbol{\gamma}_{k, i} \). If we instead employ~\eqref{eq:adapt}--\eqref{eq:combine}, we obtain an ATC-type IIR filter.
\end{example}}

\noindent {\bf Imperfect and Noisy Communication:} In the exposition so far, we have assumed for simplicity that the inter-agent communication is perfect. In practice, we may have random packet dropouts or link failures and distortions in the data exchanged by agents due to channel noise, quantization or other forms of compression. There has been extensive research on consensus and diffusion procedures that deal with time-varying or stochastic Laplacian matrices to model issues such as link failures, whereas, in other instances, controlled randomization in the communication has been used via random Laplacians as a tool to improve communication efficiency, see~\iftoggle{short}{\cite{Tsitsiklis86, Boyd-GossipInfTheory, Kar09, Zhao15}}{\cite{Tsitsiklis84, Tsitsiklis86, Xiao04, Boyd-GossipInfTheory, Kar09, Zhao15}} for a sample of the relevant literature. On the other hand, noise in the observations or communication, either injected as additive communication noise or through quantization and other forms of compression  are handled by carefully designing the mixing parameters, the $a_{\ell k}$'s in~\eqref{eq:consensus+innovation}-\eqref{eq:combine}, and building on tools from stochastic approximation as in\iftoggle{short}{ Section~\ref{sec:stochastic_approximations}}{~\eqref{eq:construction_minibatch}--\eqref{eq:construction_noisy}}~\cite{Kar12, Zhao15} or through the use of probabilistic ideas such as dithering \cite{karmoura-quantized}. Most of the development in the current article will continue to hold for such imperfect inter-agent communication through appropriate modifications as discussed above.

\nottoggle{short}{\noindent {\bf Step-sizes and convergence:} In~\eqref{eq:consensus+innovation}-\eqref{eq:combine}, the $\mu$ parameter may be viewed as a step-size or learning rate. Both the consensus + innovation and diffusion type of updates have been studied with different choices of step-sizes leading to different trade-offs. For instance, diminishing step-sizes, i.e., where the $\mu$ decays over time, may be employed to achieve convergence in an almost sure sense to an exact minimizer of the risk under broad conditions such as strongly convex risk functions (see also the discussion following~\eqref{eq:penalized_general} and connections to the penalty parameter) with strong statistical quantifications such as asymptotic normality of the iterates~\cite{KM-JSTSP, Towfic16}. A constant or non-decaying sequence of step-sizes, on the other hand, may facilitate faster mean-squared sense convergence to a neighborhood of the exact minimizer and promote adaptation (i.e., informally, the updates never stop and are able to track changes in the stochastic optimization objective) while leading to a residual error. More recently, other variants of these procedures that employ techniques such as variance reduction and gradient tracking have been employed~\cite{Yuan19, Xin20} (see also discussion below) that aim to combine the best of both worlds at the cost of more complex distributed update rules.}

\noindent \textbf{Primal-dual approaches}: As an alternative to penalty-based approaches, one may wish to enforce exact consensus by introducing constraints, such as~\cite{Sayed22}:
\vspace{-1mm}\begin{align}\label{eq:primal_dual_general}
  \min_{\{ w_k \}_{k=1}^K}\sum_{k=1}^K J_k(w_k) \ \ \ \textrm{s.t.}\ \sum_{k=1}^K \sum_{\ell \in \mathcal{N}_k} c_{\ell k} \|w_k - w_{\ell}\|^2 = 0
\end{align}
In contrast to penalty-based formulations, constrained formulations of the consensus problem can no longer be pursued using pure gradient-based algorithms. Instead, constraints are most commonly enforced through dual arguments such as ADMM, dual averaging or the augmented Lagrangian. Early algorithms involving primal-dual arguments for exact consensus optimization, such as~\iftoggle{short}{\cite{Boyd11, Towfic15, Jakovetic15, Ling15}}{\cite{Ribeiro06, Boyd11, Duchi12, Mota13:D-AMMM, Ling14:DecentralizedDynamic, Towfic15, Shi14, Jakovetic15, Koppel15, Ling15}}, involve the propagation and communication of dual variables in addition to weight vectors \( w_k \). \textcolor{black}{ADMM-based algorithms~\iftoggle{short}{\cite{Boyd11, Jakovetic15}}{\cite{Boyd11, Mota13:D-AMMM, Ling14:DecentralizedDynamic, Shi14, Jakovetic15}} generally involve \emph{two timescales}, where an auxiliary optimization problem is solved in between every outer iteration. While these methods exhibit appealing convergence properties, their implementation is only practical in situations where the inner optimization problem has a specific structure that allows it to be solved efficiently or in closed-form.}

\textcolor{black}{Single time-scale primal-dual algorithms~\iftoggle{short}{\cite{Towfic15, Ling15}}{\cite{Ribeiro06, Towfic15, Koppel15, Ling15}} instead employ first-order approxiations at every step thus avoiding the need to solve a costly inner optimization problem.} As a representative example, we list here the stochastic first-order augmented Lagrangian strategy from~\cite{Towfic15}:
\vspace{-1mm}\begin{align}
  \bcw_i =&\: {\left(I - \mu \eta \mathcal{L} \right)} \bcw_{i-1} - \mu \widehat{\nabla \mathcal{J}}(\bcw_{i-1}) - \mu \eta \mathcal{B}^{\mathsf{T}} \boldsymbol{\lambda}_{i-1} \label{eq:al_primal} \\
  \boldsymbol{\lambda}_i =&\: \boldsymbol{\lambda}_{i-1} + \mu \eta \mathcal{B} \bcw_{i-1}\label{eq:al_dual}
\end{align}
where \( L = B^{\mathsf{T}} B \). Examination of~\eqref{eq:al_primal} reveals that the augmented Lagrangian based strategy corrects the ``consensus + innovation'' algorithm~\eqref{eq:consensus+innovation} by adding the additional term \( - \mu \eta \mathcal{B}^{\mathsf{T}} \boldsymbol{\lambda}_{i-1} \), which compensates the bias induced by the penalty-based formulation~\eqref{eq:penalized_general}. While effective at ensuring exact consensus, the propagation of dual variables is associated with additional overhead in terms of both computation and communication. Conveniently, dual variables can frequently be eliminated and replaced by a momentum-like term. To illustrate this point, let us consider a variant of \eqref{eq:al_primal}--\eqref{eq:al_dual}, where the primal and dual updates are performed in an incremental manner, allowing the dual update to make use of the most recent primal variable \( \boldsymbol{\bcw}_i \), rather than \( \boldsymbol{\bcw}_{i-1} \). This results in:
\vspace{-1mm}\begin{align}
  \bcw_i =&\: {\left(I - \mu \eta \mathcal{L} \right)} \bcw_{i-1} - \mu \widehat{\nabla \mathcal{J}}(\bcw_{i-1}) - \mu \eta \mathcal{B}^{\mathsf{T}} \boldsymbol{\lambda}_{i-1} \label{eq:al_primal_inc} \\
  \boldsymbol{\lambda}_i =&\: \boldsymbol{\lambda}_{i-1} + \mu \eta \mathcal{B} \bcw_{i}\label{eq:al_dual_inc}
\end{align}
After setting \( \eta = \frac{1}{\mu} \) and following~\cite{Mokhtari16}, we can verify that~\eqref{eq:al_primal_inc}--\eqref{eq:al_dual_inc} is equivalent to:
\vspace{-1mm}\begin{align}
  \bcw_i = 2 \left( I - \mathcal{L} \right) \bcw_{i-1} - \left( I - \mathcal{L} \right) \bcw_{i-2} - \mu \left( \widehat{\nabla \mathcal{J}}(\bcw_{i-1}) - \widehat{\nabla \mathcal{J}}(\bcw_{i-2}) \right)\label{eq:EXTRA}
\end{align}
which is equivalent to the EXTRA algorithm of~\cite{Shi15} for appropriately chosen weight matrices. Letting \( A \triangleq I - L \) and returning to node-level quantities, we obtain:
\vspace{-1mm}\begin{align}
  \boldsymbol{\phi}_{k, i} =&\: \sum_{\ell \in \mathcal{N}_k} a_{\ell k} \w_{\ell, i-1} - \mu \widehat{\nabla J}_k(\w_{k, i-1}) \\
  \w_{k, i} =&\: \boldsymbol{\phi}_{k, i} + \sum_{\ell \in \mathcal{N}_k} a_{\ell k} \w_{\ell, i-1} - \boldsymbol{\phi}_{k, i-1}
\end{align}
These recursions can again be identified as a bias-corrected version of the ``consensus + innovation'' recursion~\eqref{eq:consensus+innovation}, but now rely on the momentum term \(\sum_{\ell \in \mathcal{N}_k} a_{\ell k} \w_{\ell, i-1} - \boldsymbol{\phi}_{k, i-1} \) rather than the propagation of dual variables as in~\eqref{eq:al_dual}. Making the same incremental gradient adjustment that led to the penalty-based ATC-diffusion algorithm~\eqref{eq:adapt}--\eqref{eq:combine}, we obtain the Exact diffusion algorithm from~\cite{Yuan18}:
\vspace{-1mm}\begin{align}
 \boldsymbol{\psi}_{k, i} =&\: \w_{k, i-1} - \mu \widehat{\nabla J}_k(\w_{k, i-1}) \\
  \boldsymbol{\phi}_{k, i} =&\: \boldsymbol{\psi}_{k, i} + \w_{k, i-1} - \boldsymbol{\psi}_{k, i-1} \\
  \w_{k, i} =&\: \sum_{\ell \in \mathcal{N}_k} a_{\ell k} \boldsymbol{\phi}_{\ell, i-1} 
\end{align}
Exact diffusion is also referred to as $D^2$~\cite{Tang18} or NIDS\nottoggle{short}{~\cite{Li19}}{}.

\noindent \textbf{Gradient-tracking based approaches:} An alternative to the approaches described above is based on gradient tracking. While the initial motivation~\cite{Xu15, DiLorenzo16} for the construction was based on the dynamic average consensus algorithm\nottoggle{short}{~\cite{Kia19}}, it has been noted in~\cite{Nedic17} that gradient-tracking-based algorithms for decentralized optimization can be viewed as a variation of the primal-dual arguments leading to the EXTRA and Exact diffusion algorithms described above. We refer the interested reader to~\cite{Nedic17, Sayed22} for details, and simply list the resulting NEXT~\cite{DiLorenzo16}/DIGing~\cite{Nedic17} algorithm:
\vspace{-1mm}\begin{align}
  \boldsymbol{\phi}_{k, i} =&\: \sum_{\ell \in \mathcal{N}_k} a_{\ell k} \w_{\ell, i-1} - \mu \boldsymbol{g}_{k, i-1} \\
  \boldsymbol{g}_{k, i} =&\: \sum_{\ell \in \mathcal{N}_k} a_{\ell k} \boldsymbol{g}_{\ell, i} + \widehat{\nabla J}_k(\w_{k, i-1}) - \widehat{\nabla J}_k(\w_{k, i-2})
\end{align}
Following an incremental construction on the other hand, analogously to the step from EXTRA to Exact diffusion before, results in an ATC-variant of the NEXT/DIGing algorithm, proposed in~\cite{Xu15} under the name Aug-DGM:
\vspace{-1mm}\begin{align}
  \boldsymbol{\psi}_{k, i} =&\: \w_{k, i-1} - \mu \boldsymbol{g}_{k, i-1} \\
  \boldsymbol{\phi}_{k, i} =&\: \sum_{\ell \in \mathcal{N}_k} a_{\ell k} \boldsymbol{\psi}_{\ell, i-1} \\
  \boldsymbol{g}_{k, i} =&\: \sum_{\ell \in \mathcal{N}_k} a_{\ell k} \left( \boldsymbol{g}_{\ell, i} + \widehat{\nabla J}_k(\w_{\ell, i-1}) - \widehat{\nabla J}_k(\w_{\ell, i-2}) \right)
\end{align}
Decentralized algorithms for consensus optimization based on gradient-tracking generally have similar convergence properties to their primal-dual counterparts EXTRA and Exact diffusion. One key difference is the fact that exchanges of the gradient estimate \( \boldsymbol{g}_{k, i} \) in addition to the local models results in an increase in communication cost roughly at a factor of two.

\noindent \textcolor{black}{\textbf{Constrained Learning:} In our discussion so far, the constraint \(\Omega\) of~\eqref{eq:generic_couple} has been used to encode equality constraints of the form \( \Omega = \{ \cw | \sum_{k=1}^K \sum_{\ell \in \mathcal{N}_k} c_{\ell k} \|w_k - w_{\ell}\|^2 = 0 \} = \{ \cw | w_{k} = w \ \forall \ k\}\). This ensures consensus on a common model \( w\). In many application, we may wish to further constrain the common model \( w \) to some set \( \Theta \). Variations of most algorithms described in Sec.~\ref{sec:unifiedview} for constrained optimization and learning have been developed and studied by employing Euclidean and proximal projections or penalty functions~\iftoggle{short}{\cite{Ram10, Shi15:AProximalGradient, Towfic15-2014,rec_nonlin_est:Sahu}}{\cite{Ram10, Shi15:AProximalGradient, Vlaski15, Towfic15-2014,rec_nonlin_est:Sahu, Alghunaim21:DecentralizedProximalGradient}}. These solve the constrained consensus optimization problem\iftoggle{short}{ \( \min_{ w \in \Theta }\sum_{k=1}^K J_k(w) \).}{:
\vspace{-1mm}\begin{align}\label{eq:constrainted_consensus_optimization_problem}
  \min_{ w \in \Theta }\sum_{k=1}^K J_k(w)
\end{align}}
For example, applying the same incremental argument that led to~\eqref{eq:adapt}--\eqref{eq:combine}, followed by projection onto \( \Theta \), leads to a projected variant of the ATC-diffusion or consensus + innovation algorithm~\eqref{eq:consensus+innovation}--\eqref{eq:combine}, studied in~\cite{Ram10,rec_nonlin_est:Sahu}:
\vspace{-1mm}\begin{align}
  \boldsymbol{\psi}_{k, i} =&\: \w_{k, i} - \mu \widehat{\nabla J}_k(\w_{k, i-1})\label{eq:projected_adapt} \\
  \w_{k, i} =&\: \mathrm{Proj}_{\Theta} \left( \sum_{\ell \in \mathcal{N}_k} {a}_{\ell k} \boldsymbol{\psi}_{\ell, i} \right)\label{eq:projected_combine}
\end{align}
Similarly, we may introduce projections into primal-dual algorithms to derive projected variants of primal-dual algorithms, such as the PG-EXTRA generalization~\cite{Shi15:AProximalGradient} of the EXTRA algorithm~\eqref{eq:EXTRA}\nottoggle{short}{ or general primal-dual algorithms~\cite{Alghunaim21:DecentralizedProximalGradient}}.}

\noindent \textbf{Multitask Learning:} While the pursuit of an optimal average model as defined in~\eqref{eq:consensus_optimization_problem} is appropriate in many situations, it is important to recognize that a good average model may perform poorly on any local cost \( J_k(\cdot) \). This observation motivates the pursuit of networked multitask learning algorithms~\cite{Nassif20}, where agents aim to learn from one another without forcing exact consensus. More recently, this area has received attention under the name of personalized federated learning\nottoggle{short}{~\cite{Kairouz21}}. Multitask learning is generally achieved using variations of the regularized aggregate problem~\eqref{eq:generic_couple}, where the regularization is chosen to match some underlying prior on task relationships (rather than to enforce exact consensus). Solutions can again be pursued using primal or primal-dual approaches. Given space limitations, a detailed treatment of multitask learning is beyond the scope of this manuscript, and we refer the reader to~\cite{Nassif20} and the references therein.

\vspace{-7mm}\textcolor{black}{\subsection{Applications}\label{sec:applications}
\subsubsection{Weather Prediction}
The task of predicting weather patterns naturally lends itself to networked solutions because \( (a) \) measurements tend to be available in dispersed locations and \( (b) \) it is reasonable to believe that weather models ought to be related in adjacent regions, encouraging the diffusion of information as a means of improving performance. To illustrate this fact, we reproduce here a simulation study from~\cite{Nassif20:LearningOverMultitaskGraphs}. The simulation is based on meteorological data from across the United States, shown in the top panel of Fig.~\ref{fig:weather}. The implementation is based on the regularized learning problem~\eqref{eq:mse_example} with logistic risks \( J_k(w_k) = \mathds{E} \ln\left( 1 + e^{- \boldsymbol{\gamma}_{k, i} \boldsymbol{h}_{k, i}^{\mathsf{T}} w_k} \right) + \rho \|w_k\|^2 \). Performance is shown in the bottom panel of Fig.~\ref{fig:weather}, where the choice \( \eta = 0\) corresponds to a non-cooperative implementation, \( \eta = \mu^{-1} \) corresponds to the ATC-diffusion algorithm~\eqref{eq:adapt}--\eqref{eq:combine}, and other choices of \( \eta \) correspond to softer coupling of local models. Due to space limitations, we refer the reader to~\cite{Nassif20:LearningOverMultitaskGraphs} for a more detailed discussion of the setup and results.
\begin{figure}[htb]
  \centering
	\includegraphics[width=.8\columnwidth]{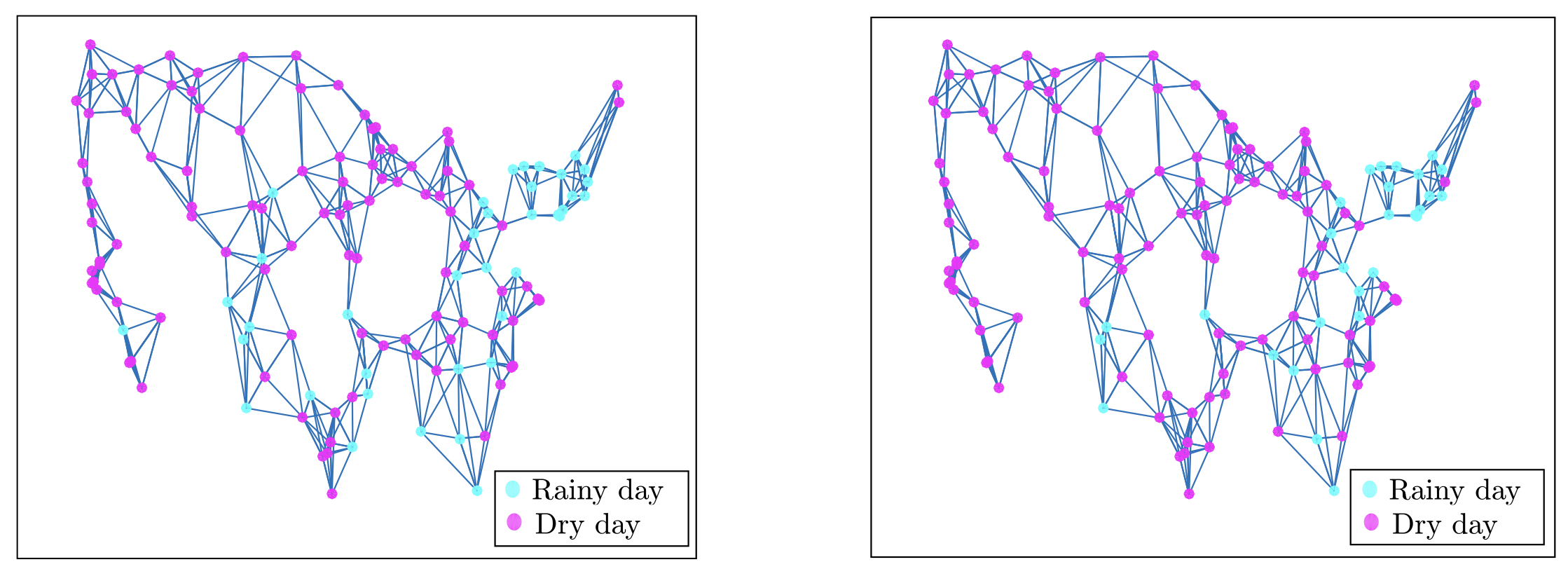}
 \includegraphics[width=.8\columnwidth]{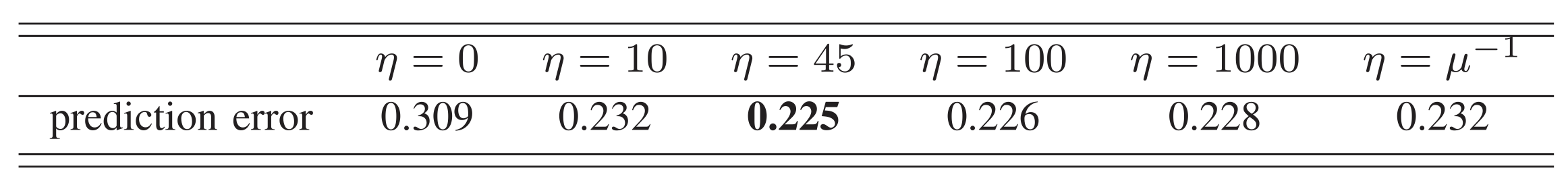}
\caption{\small Weather prediction using diffusion algorithms (taken form~\cite{Nassif20:LearningOverMultitaskGraphs}, CC-BY License). (\emph{Top left}) Actual occurrence of rain. (\emph{Top right}) Predicted occurrence of rain. (\emph{Bottom}) Prediction accuracy as a function of the regularization parameter \( \eta \) of \eqref{eq:mse_example}.}\label{fig:weather}
\end{figure}
\subsubsection{Wide-Area Monitoring in Power Systems}
Wide-area monitoring in power transmission systems consists of tracking the overall system state based on measurements obtained at control areas or balancing authorities (nodes or agents in our current exposition). The geographical distribution and practical data sharing limitations among the control areas naturally calls for distributed state estimation algorithms (see~\cite{wide_area_tsg:Xie}) with the goal of monitoring the global system state while minimizing data exchange among the control areas. In~\cite{wide_area_tsg:Xie} fully distributed approaches for wide-area state estimation based on consensus + innovation type algorithms (see~\eqref{eq:consensus+innovation}) are proposed, both for DC and AC state estimation. The typical quantities of interest in wide-area monitoring are voltage magnitudes and relative angles (phases) at the system buses based on power flow measurements at subsets of transmission lines and power injection measurements at the system buses. In DC state estimation the bus voltage magnitudes are typically assumed to be at a nominal 1.0 p. u. reference value (see~\cite{wide_area_tsg:Xie} for details) and the unknown phase estimation at the buses reduces to a linear least squares type formulation as in~\eqref{eqn:linearobsmodel}. In particular, Fig.~\ref{fig:IEEE_118} shows an application of a consensus + innovation approach with decaying step sizes (taken from~\cite{wide_area_tsg:Xie}) for DC state estimation on an IEEE 118-bus benchmark test system: Fig.~\ref{fig:IEEE_118} (on the left) depicts the 118-bus system partitioned into 8 control areas that communicate over a connected communication graph (typically this graph conforms to the physical coupling between the control areas or is chosen based on geographical proximity); the application essentially consists of reformulating the wide-area phase estimation objective as a least-squares cost minimization with $J_{k}(w_{k})$ of the form in~\eqref{eqn:linearobsmodel}-\eqref{eq:lin_obs_cost_model} and applying the consensus + innovation approach. In Fig.~\ref{fig:IEEE_118} (on the right) we compare the gap (referred to as the phase angle gap) between the relative phases obtained by the iterative distributed approach and those from a hypothetical fusion center based optimal one-shot least-squares estimator across multiple bus pairs, i.e., for instance, the quantity $g_{1,2}$ denotes the gap between the phase difference between buses 1 and 2 obtained by the distributed approach and that obtained by the centralized estimator. As expected, by the convergence guarantees discussed in Section~\ref{sec:penalty_based}, these gaps converge to zero as the iterations progress. Similar distributed approaches may be used for distributed (nonlinear) AC state estimation where the objective is to estimate both the bus voltages and relative angles. This is performed by resorting to a nonlinear least squares type minimization in~\cite{wide_area_tsg:Xie, rec_nonlin_est:Sahu} and applying a projected variant of the consensus + innovation approach (see the discussion pertaining to~\eqref{eq:projected_adapt}--\eqref{eq:projected_combine}) to deal with certain trigonometric nonlinearities associated with the AC power flow model.
\begin{figure}[htb]
  \centering
	\includegraphics[width=.47\columnwidth]{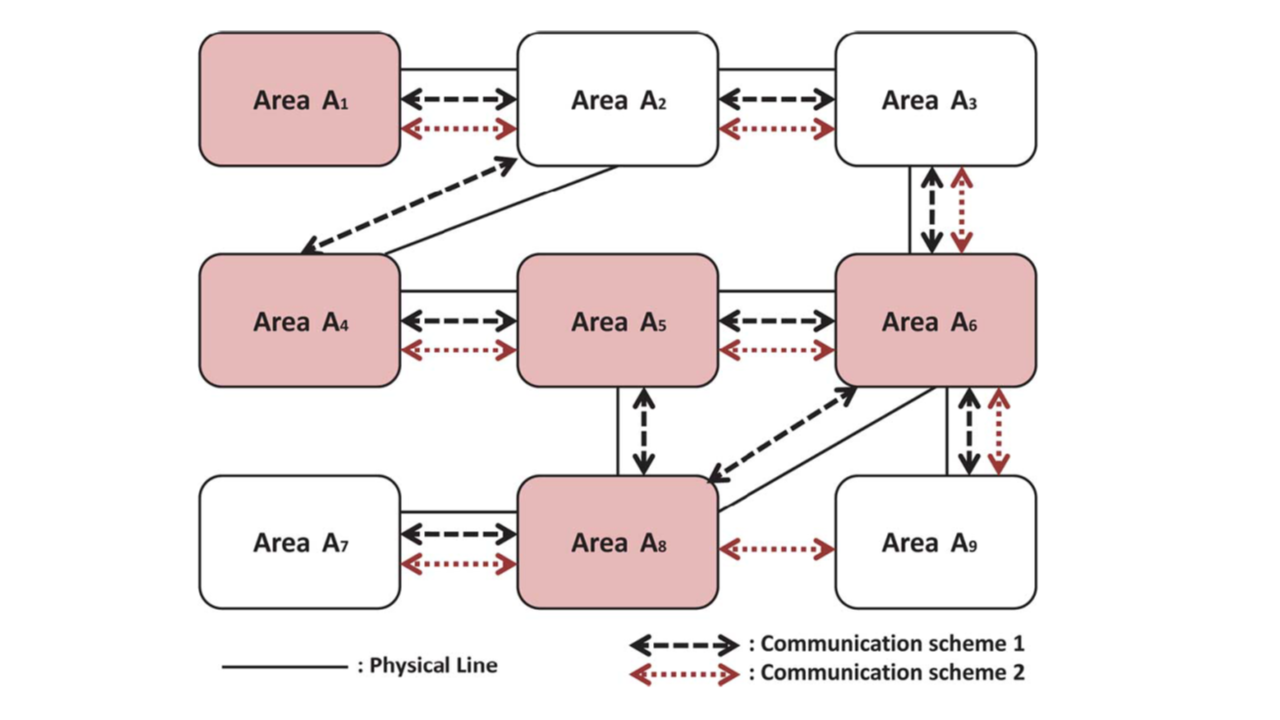}
 \includegraphics[width=.5\columnwidth]{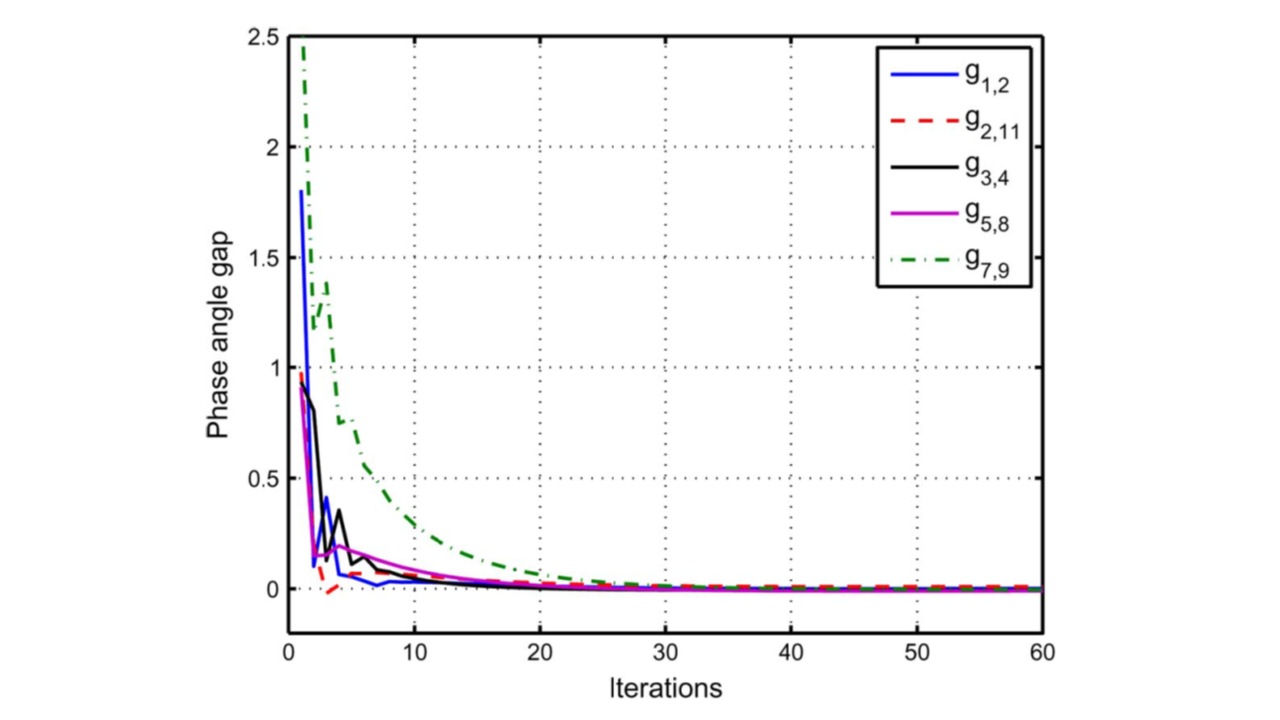}
\caption{\small Wide-area state estimation using consensus + innovation algorithms (taken form~\cite{wide_area_tsg:Xie}, CC-BY License). (\emph{Left}) IEEE 118-bus system partioned into 8 control areas or nodes with possible inter-node communication patterns. (\emph{Right}) Relative phase angle estimation error at different bus pairs.}\label{fig:IEEE_118}
\end{figure}}

\vspace{-4mm}\section{Performance Guarantees in Distributed Learning}\vspace{-2mm}
We now proceed to survey some performance guarantees of algorithms for decentralized learning, with a particular emphasis on stochastic settings. Given space limitations, it is not possible for us to provide a comprehensive survey. Instead, we aim to highlight some key insights that have emerged from an extensive body of work over the past two decades, in an attempt to provide the reader with a starting point and guidelines when matching the choice of a learning algorithm to a problem at hand.

\nottoggle{short}{Before proceeding we briefly review and relate some of the key performance metrics that appear across different communities in the literature. Due to lack of space, we focus here on optimization metrics and  refer the reader to the literature when noise and random failures occur in sensing and communication. When this is the case, large deviations show consistency and efficiency of distributed detectors and estimators under broad conditions \cite{Kar12,bajovic2012large, Matta16}. With regard to optimization metrics, the \emph{mean-squared deviation} (MSD) of an iterate \( \w_{k, i} \) relative to some limiting point \( w^o \) is defined as \( \mathrm{MSD}_{k, i} \triangleq \E \|w^o - \w_{k, i} \|^2 \) and measures the expected squared Euclidean distance to the target \( w^o \). The \emph{excess-risk} (ER) on the other hand measures the sub-optimality \( \mathrm{ER}_{k, i} \triangleq \mathds{E} J(\w_{k, i}) - J(w^o) \) directly. In many cases, depending on regularity conditions of the cost \( J(\cdot) \), MSD and ER can be related and inferred from one another. For example, for \( \delta \)-smooth and \( \nu \)-strongly-convex \( J(\cdot) \) it holds that \( \frac{\nu}{2} \mathrm{MSD}_{k, i} \le \mathrm{ER}_{k, i} \le \frac{\delta}{2} \mathrm{MSD}_{k, i} \). Another measure of performance, more commonly employed in the context of online optimization (see, e.g.,~\cite{Koppel15}) is the \emph{regret} defined as \( \mathrm{Regret}_{k, i} \triangleq \sum_{j=1}^i \mathrm{ER}_{k, j} \). In other words, we may view the regret of an algorithm as the cumulative excess risk. The final commonly employed performance metric is the \emph{iteration complexity} (IR), which measures the number of iterations required to achieve a given excess-risk \( \epsilon \), i.e., \( \mathrm{IR}_k \triangleq \min_{i} \mathrm{ER}_{k, i} \ \mathrm{s.t.}\ \mathrm{ER}_i \le \epsilon \). We note that both the regret and iteration complexity of an algorithm can be inferred directly from a complete characterization of the excess-risk or mean-squared deviation, while providing a more granular characterization of the transient behavior of the algorithm. As such, we will report on the MSD, and ER whenever available with the understanding that regret and iteration complexity can be recovered as needed. For each of these metrics, we may define the average across the network and omit the subscript \( k \). In this way for example, \( \mathrm{MSD}_i \triangleq \frac{1}{K} \mathrm{MSD}_{k, i} \).

MSD, ER, regret and IR are all useful measures of performance particularly for \emph{convex} optimization problems, where unique optimal solutions \( w^o \) can be defined and pursued efficiently by stochastic gradient methods. The pursuit of performance guarantees for nonconvex optimization and learning, driven in large part by an effort to understand and certify the empirical success of deep learning algorithms, has motivated a need to develop optimality measures for nonconvex optimization. One measure of optimality commonly employed for nonconvex optimization is first-order stationarity, i.e., the squared expected norm of the gradient \( \mathds{E} \|\nabla J(\w_{k, i})\|^2 \). While this measure is useful, since every local minimum of a cost function must necessarily have small local gradient norm, first-order stationary points may include saddle-points, which are known to pose bottlenecks in many nonconvex learning problems. To exclude strict saddle-points from the set of limiting points, we may instead study the iteration complexity of finding points that are jointly first-order stationary (i.e., \( \|\nabla J(\w_{k, i})\|^2 \le \epsilon \)) and exhibit a Hessian that is bounded from below, \( \nabla^2 J(\w_{k, i}) \ge -\tau \). Such points are referred to as second-order stationary and correspond to ``good'' solutions for many nonconvex learning problems. We refer the reader to~\iftoggle{short}{\cite{Vlaski19nonconvexP2}}{\cite{Ge15, Vlaski19nonconvexP2}} and the references therein for a more detailed discussion.}

\vspace{-4mm}\subsection{Constant and Diminishing Step-Sizes}\vspace{-2mm}\label{eq:constant_and_diminishing}
Optimization algorithms based on stochastic gradient approximations are subject to persistent noise, and hence generally do not converge to exact solutions. This can be remedied by employing a time-varying and diminishing step-size, resulting in slower but exact convergence. We highlight this by comparing the performance guarantees of the primal consensus and diffusion algorithms from~\iftoggle{short}{\cite{Kar12, Lopes08, Chen15transient}}{\cite{Kar12, Lopes08, Chen13, Chen15transient}}, though similar conclusions apply to other decentralized algorithms. For strongly-convex costs and using a constant step-size construction, the asymptotic performance of the penalty-based algorithms described in Sec.~\ref{sec:penalty_based} is given by~\cite[Example 11.8]{Sayed14}:
\vspace{-1mm}\begin{align}\label{eq:er_constant}
  \limsup_{i \to \infty} \mathrm{ER}_i = \frac{\mu}{4} \sum_{k=1}^K p_k^2 \sigma_k^2 + o(\mu)
\end{align}
\iftoggle{short}{The \emph{excess-risk} (ER) measures the average sub-optimality \( \mathrm{ER}_{i} \triangleq \frac{1}{K} \sum_{k=1}^K \mathds{E} J(\w_{k, i}) - J(w^o) \). }{}\iftoggle{short}{The notation}{Here,} \( o(\mu) \) denotes terms which are higher-order in the step-size and hence negligible for sufficiently small step-sizes \( \mu \). \textcolor{black}{The constants \( \sigma_k^2 \) correspond to the absolute gradient noise variance of~\eqref{eq:variance_bound_error}}. The analysis in~\cite{Sayed14} is performed for general left-stochastic adjacency matrices \( A \neq A^{\mathsf{T}} \) with Perron eigenvector \( A p = p \). For symmetric adjacency matrices, the weights reduce to \( p_k = \frac{1}{K} \). Convergence to the steady-state value occurs at a linear rate given by \( \alpha = 1 - 2 \nu \mu + o(\mu) \)~\cite[Eq.~(11.139)]{Sayed14}, where \( \nu \) denotes the strong-convexity constant of the aggregate cost \( J(w) \). If we instead employ a diminishing step-size of the form \( \mu_i = \frac{1}{i} \), it holds asymptotically for large \( i \) that~\cite{Towfic16}:
\vspace{-1mm}\begin{align}\label{eq:er_diminishing}
  \mathrm{ER}_i = \frac{\mu}{4 i} \sum_{k=1}^K p_k^2 \sigma_k^2 + o(\mu)
\end{align}
and hence \( \lim_{i \to \infty} \mathrm{ER}_i = 0 \). At the same time, we note that convergence using a diminishing step-size occurs at a sublinear rate \( O\left(\frac{1}{i}\right) \) rather than the linear rate \( O\left( (1 - 2\mu \nu + o(\mu))^i \right) \). 

\textcolor{black}{These remarks reflect a well-known trade-off between convergence rate, asymptotic error, and tracking ability of a learning algorithm \cite{Kar12,kar2013distributed,Towfic16}. Algorithms with vanishing step-sizes can converge asymptotically to the exact minimizer with zero error albeit at a slower rate than when constant step-sizes are used. In this latter case, the algorithms approach the minimizer at a faster exponential rate, albeit within an MSE range that is proportional to the step-size parameter. When this parameter is small, as is normally the case, this construction enables the algorithm with a constant step-size to track drifts in the underlying parameter when the statistical properties of the data change with time. Often times, implementations in practice may use a combination of vanishing and constant step-sizes. On the other hand, when one is interested in asymptotic convergence of the error to zero, then it is known from statistical learning theory and large-sample asymptotics in parameter estimation that the $O\left(\frac{1}{i}\right)$ rate is optimal for statistically consistent online estimators (i.e., estimators that achieve asymptotically zero error almost surely), where~$i$, the iteration count,
coincides with or is proportional to the number of data points or online stochastic gradients
sampled during the estimation process. Further, in this online scenario, within the class of
statistically consistent estimators, the ones with optimal asymptotic variance, i.e., asymptotically
efficient estimators, may be obtained by appropriately tuning the diminishing step-size
sequences (see, for example, \cite{Kar12,kar2013distributed,Towfic16}); in such scenarios the distributed estimators end up
achieving the optimal online centralized error rates.}


\textcolor{black}{Both, nonvanishing and vanishing weights algorithms, can overcome lack of knowledge of model parameters or noise statistics, for example, by replacing noise mean and covariance by empirical sample estimates, like distributed RLS \cite{Cattivelli10}, still guaranteeing the distributed algorithms stability and error mean and covariance asymptotic optimality. With vanishing weights, to guarantee optimal asymptotic mean square error in the sense of Fisher information rate, algorithm~\eqref{eq:consensus+innovation} should be augmented by a recursion for the gain of the innovations or data term \cite{kar2013distributed}, so that the agents engage  on distributed learning to recover asymptotically the optimal gains, while  simultaneously carrying out their distributed task with negligible asymptotic information rate loss.
%
}



\vspace{-4mm}\subsection{Linear Gains in Performance}\vspace{-2mm}
If we consider symmetric adjacency matrices \( A = A^{\mathsf{T}} \), resulting in \( p_k = \frac{1}{K} \), and homogeneous data profiles \( \sigma_k^2 = \sigma^2 \), we note that for both constant~\eqref{eq:er_constant} and diminishing step-sizes~\eqref{eq:er_diminishing}, the asymptotic excess risk scales with \( O\left( \frac{\mu \sigma^2}{K} \right) \). The scaling by the network size \( K \) is referred to as linear gain. It is consistent with the performance gains that can be expected when fusing raw data in a centralized architecture~(see, e.g.,~\cite[Theorem 5.1]{Sayed14}), and provides motivation for agents to participate in the cooperative learning protocol. Analogous results have been obtained for primal-dual algorithms~\cite{Yuan20}, as well as in the pursuit of \nottoggle{short}{first-order stationary~\cite{Lian17} and }second-order stationary-points~\iftoggle{short}{\cite{Vlaski19nonconvexP2}}{\cite{Vlaski19nonconvexP2, Vlaski20}} points in nonconvex environments.
\nottoggle{short}{\begin{figure}[htb]
  \centering
	\includegraphics[width=.8\columnwidth]{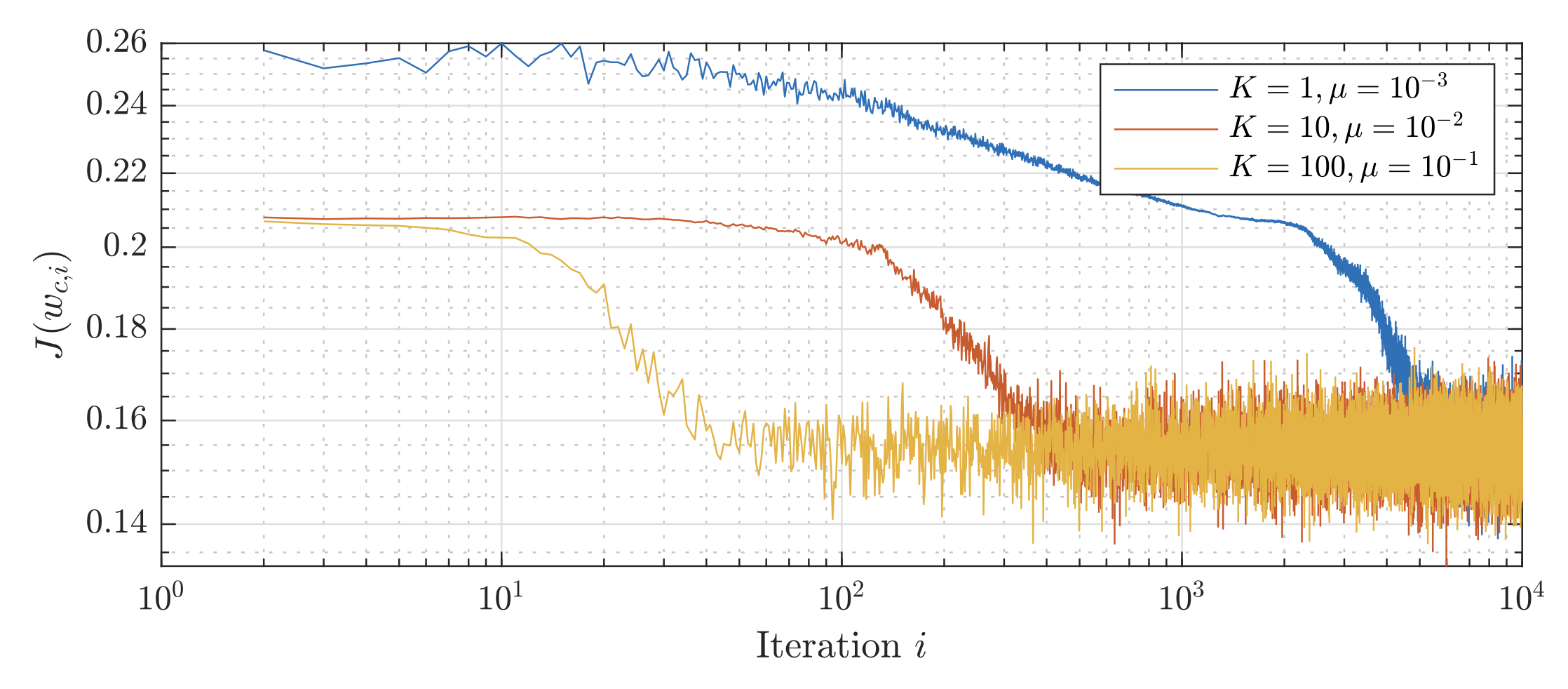}
\caption{\small \textcolor{black}{Illustration of a linear performance gain (taken form~\cite{Vlaski20}, {\textcopyright}IEEE 2020). The time it takes a network to escape from a saddle-point decays by a factor of ten as the size of the network increases by a factor of ten.}}\label{fig:linear_speedup}
\end{figure}}

\vspace{-4mm}\subsection{Penalty-Based and Primal-Dual Algorithms}\vspace{-2mm}
A motivation for considering primal-dual algorithms for decentralized optimization over penalty-based construction is the removal of the bias induced by employing a finite regularization term~\eqref{eq:penalized_general} in place of~\eqref{eq:primal_dual_general}. When exact gradients are employed and no noise is added due to the use of stochastic gradient approximations, this results in a pronounced difference in performance, as primal-dual algorithms are able to converge linearly and exactly, using a constant step-size, in strongly-convex environments~\cite{Shi15, Yuan18, Jakovetic15, Ling15, Nedic17}, while penalty-algorithms require a diminishing step-size to ensure exact convergence, resulting in a sublinear rate~\cite{Nedic09}.

In the stochastic setting, however, iterates are subjected to additional perturbations induced by the utilization of data-dependent, stochastic gradient approximations. This causes the difference in performance between penalty-based and primal-dual algorithms to be more nuanced. For example, it was shown in~\cite{Towfic15} that the primal-dual algorithm~\eqref{eq:al_primal}--\eqref{eq:al_dual} exhibits strictly worse performance than penalty-based approaches such as consensus and diffusion algorithms, when constant step-sizes and stochastic gradient approximations are employed. This is due to the fact that the penalty-based algorithms exhibit lower variance in steady-state, which compensates for the additional bias. On the other hand, stochastic variants of Exact diffusion~\cite{Yuan18} and gradient-tracking~\cite{DiLorenzo16} have been shown analytically and empirically to improve upon the performance of their penalty-based counterparts. We illustrate this by reviewing the results of~\cite{Yuan20} as a case study. For the diffusion algorithms as an example of a penalty-based algorithm, the authors derived a refined version of the bound~\eqref{eq:er_constant} of the form:
\vspace{-1mm}\begin{align}\label{eq:msd_diffusion_precise}
  \limsup_{i \to \infty} \textrm{MSD}_i = O\left( \frac{\mu \sigma^2}{K} + \frac{\mu^2 \lambda^2 \sigma^2}{1-\lambda} + \frac{\mu^2 \sigma^2 b^2}{(1-\lambda)^2} \right)
\end{align}
\iftoggle{short}{The \emph{mean-squared deviation} (MSD) of the network is defined as \( \mathrm{MSD}_{i} \triangleq \frac{1}{K} \sum_{k=1}^K \E \|w^o - \w_{k, i} \|^2 \). For  \( \delta \)-smooth and \( \nu \)-strongly-convex \( J(\cdot) \) it holds that \( \frac{\nu}{2} \mathrm{MSD}_{i} \le \mathrm{ER}_{i} \le \frac{\delta}{2} \mathrm{MSD}_{i} \).}{} The first term in the performance expression \( \frac{\mu \sigma^2}{K} \) corresponds to the performance deterioration from employing stochastic gradient approximations with variance \( \sigma^2 \), and is proportional to the step-size \( \mu \). This term is consistent with~\eqref{eq:er_constant}. The other two terms scale with \( \mu^2 \) and quantify the interplay between the mixing rate of the adjacency matrix \( \lambda = \rho\left( A - \frac{1}{K} \mathds{1}\mathds{1}^{\mathsf{T}} \right) \) and the bias term \( b^2 = \frac{1}{K} \sum_{k=1}^K \|\nabla J_k(w^o)\|^2 \). The mixing rate \( \lambda \) measures the level of connectivity of the network, and is close to one whenever the adjacency matrix is sparse. The bias term \( b^2 \) on the other hand measures the level of heterogeneity in the network. Both \( O(\mu^2) \)-terms become negligible as \( \mu \to \infty \), but can be significant for very sparse, heterogeneous networks and moderate step-sizes.

For Exact diffusion, as an example of a stochastic primal-dual algorithm, on the other hand, we have~\cite{Yuan18, Yuan20}:
\vspace{-1mm}\begin{align}
  \limsup_{i \to \infty} \textrm{MSD}_i = O\left( \frac{\mu \sigma^2}{K} + \frac{\mu^2 \lambda^2 \sigma^2}{1-\lambda} \right)\label{eq:msd_exact_diffusion}
\end{align}
We note the removal of the term \( \frac{\mu^2 \sigma^2 b^2}{(1-\lambda)^2} \). As a result, the performance no longer depends on the heterogeneity \( b^2 \), and has an improved dependence on the mixing rate \( \lambda \). Similar improved dependence on network heterogeneity and connectivity has been observed in pursuing first-order stationary points of nonconvex problems as well~\cite{Tang18}.
\begin{figure}[htb]
  \centering
	\includegraphics[width=\columnwidth]{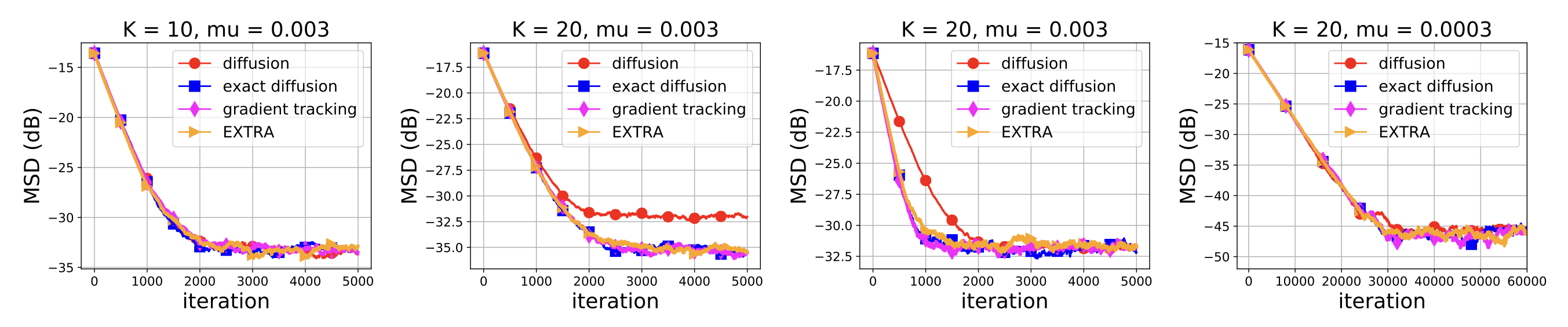}
\caption{\small \textcolor{black}{Illustration of the benefit of primal-dual algorithms (taken form~\cite{Yuan20}, {\textcopyright}IEEE 2020). Primal-dual algorithms (EXTRA, gradient-tracking, exact diffusion) outperform a primal algorithm (diffusion) for a large network size (large \( K \), i.e., \( \lambda \) closer to 1) and large step-size. This is the range where \( \frac{\mu^2 \sigma^2 b^2}{(1-\lambda)^2} \) in~\eqref{eq:msd_diffusion_precise} is non-negligible.}}\label{fig:primal_dual}
\end{figure}

\vspace{-4mm}\subsection{Stability Gains via Incremental Constructions}\vspace{-2mm}
As we saw throughout the algorithm derivations in Sec~\ref{sec:derivations}, the derivations of some decentralized algorithms rely on the use of incremental steps, such as the ATC-diffusion algorithm~\cite{Lopes08}, the Exact diffusion algorithm~\cite{Yuan18}, and the Aug-DGM algorithm~\cite{Xu15}. These variants incorporate incremental steps in comparison to the ``consensus+innovation'' algorithm~\cite{Nedic09, Kar12}, the EXTRA algorithm~\cite{Shi15}, and the DIGing algorithm~\cite{Nedic17}. It turns out that in many cases the incremental steps endow the resulting algorithms with improved robustness and stability properties, particularly when employing constant and uncoordinated step-sizes and noisy gradient approximations. Early evidence to this phenomenon appears in~\iftoggle{short}{\cite{Sayed14, Sayed14proc}}{\cite{Tu12, Sayed14, Sayed14proc}}, where it was shown that diffusion strategies based on the adapt-then-combine (ATC) construction, which is incremental, enjoy a wider stability range than consensus based constructions. In particular, the stability range for the ATC-diffusion algorithm can be independent of the network connectivity (so long as agents are locally stable), while that for the consensus algorithm in general depends on the mixing rate of the adjacency matrix~\iftoggle{short}{\cite{Sayed14, Sayed14proc}}{\cite{Tu12, Sayed14, Sayed14proc}}. Analogous observations were made in~\cite{Yuan18} when comparing the stability range of the Exact diffusion algorithm to EXTRA.

\vspace{-4mm}\subsection{Asynchronous Behavior}\vspace{-2mm}
The stochastic gradient approximations framework described in Sec.~\ref{sec:stochastic_approximations} is general enough to cover a large number of phenomena that may arise in the presence of asynchrony and imperfections, such as intermittent updates~\eqref{eq:construction_async} or noisy links\nottoggle{short}{~\eqref{eq:construction_noisy}}. The implications of these imperfections on performance follow from relations~\eqref{eq:er_constant}--\eqref{eq:msd_exact_diffusion} after adjusting the gradient variance \( \sigma_k^2 \) according to Table~\ref{tab:variations}. Another form of asynchrony, not directly covered within the gradient approximations framework, refers to time-varying, intermittent communication graphs. Such asynchrony can be more challenging, since exchanges that now occur so infrequently can in principle result in divergent behavior, particularly for heterogeneous networks. Nevertheless, when properly designed, decentralized algorithms have been shown to be remarkably robust to asynchronous communication policies, including random~\cite{Zhao15} and deterministically time-varying policies~\cite{Nedic17}. The take-away from these studies is that, as long as adjacency matrices are connected in expectation~\cite{Zhao15}, or their union over time is connected~\cite{Nedic17}, information can sufficiently diffuse, and agents can efficiently learn from each other.

\vspace{-4mm}\subsection{Asymmetric Combination Policies}\vspace{-2mm}\label{sec:asymmetric}
Most of our discussion so far has focused on symmetric adjacency matrices \( A = A^{\mathsf{T}} \). Nevertheless, decentralized algorithm for optimization and learning can also be deployed with asymmetric matrices~\iftoggle{short}{\cite{Lopes08, Chen15transient, Sayed14, Yuan18, Vlaski19nonconvexP2}}{\cite{Lopes08, Chen13, Chen15transient, Sayed14, Yuan18, Vlaski19nonconvexP2}}. The effect of such constructions is that certain agents will be able to exert more or less influence over the behavior of the network. To be precise, we associate with the adjacency matrix its Perron eigenvector \( A p = p \), where \( p_k \) denotes the entry corresponding to agent \( k \). It can then be shown that most decentralized algorithms will converge to the minimizer of the weighted sum\iftoggle{short}{, i.e., \( w^o \triangleq \sum_{k=1}^{K} p_k J_k(w) \),}{:
\vspace{-1mm}\begin{align}
  w^o \triangleq \sum_{k=1}^{K} p_k J_k(w)
\end{align}}
where the weights \( p_k \) now modulate the relative influence of the cost \( J_k(w) \) associated with agent \( k \). For symmetric matrices \( A = A^{\mathsf{T}} \), we have \( p_k = \frac{1}{K} \) and we recover~\eqref{eq:consensus_optimization_problem}. The ability for certain agents to be more or less influential within the network adds a degree of freedom to the design of multi-agent system. In heterogeneous environments, where some agents may have access to data or gradient approximations of higher quality, this can be exploited to improve performance or convergence rate~\iftoggle{short}{\cite{Sayed14}}{\cite{Sayed14, Vlaski20}}. On the other hand, there may be situations where such behavior is undesirable, and we may wish to minimize the unweighted cost~\eqref{eq:consensus_optimization_problem} while employing asymmetric network topologies. This can be achieved by effectively rescaling the agent-specific step-sizes to compensate for the Perron weights \( p_k \)~\cite{Yuan18, Nedic15, Saadatniaki20}.
\nottoggle{short}{\begin{figure}[htb]
  \centering
	\includegraphics[width=.6\columnwidth]{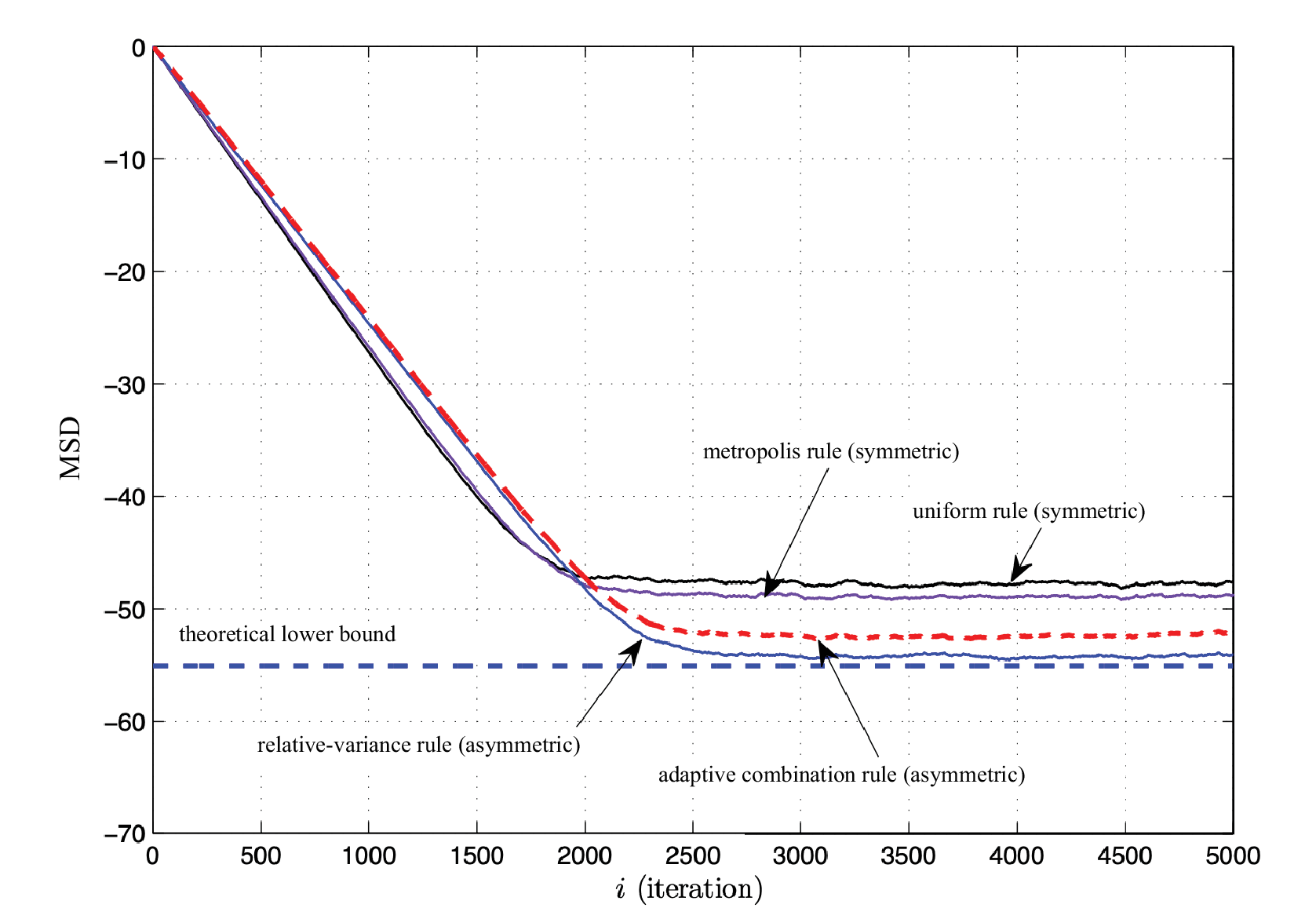}
\caption{\small \textcolor{black}{Illustration of the benefit of employing asymmetric combination policies (taken form~\cite{Sayed14}). The relative variance rule (left-stochastic) results in lower limiting MSD than symmetric policies.}}\label{fig:left_stochastic}
\end{figure}}

\vspace{-4mm}\subsection{Federated Learning}\vspace{-2mm}
Federated learning\nottoggle{short}{~\cite{Kairouz21}}{} has emerged in recent years as an umbrella term for architectures that involve a fusion center, as well as high levels of asynchrony and heterogeneity. The federated setting can be viewed as a special case of the decentralized algorithms for appropriately chosen network topologies and asynchrony models. As a result, many algorithms and performance guarantees for federated learning can be recovered from their decentralized counterparts by appropriately specializing the network topology. To illustrate this fact, let us consider the diffusion algorithm~\eqref{eq:adapt}--\eqref{eq:combine} with random adjacency matrix \( \boldsymbol{A} \). The resulting behavior at any given agent corresponds precisely to a stochastic variant of the federated averaging algorithm\nottoggle{short}{ from~\cite{Kairouz21}}, and the analysis of~\cite{Zhao15} applies. We may instead consider a deterministic variant with time-varying \( A_i \), where \( A_i = \mathds{1}\mathds{1}^{\mathsf{T}} \) if \( i \) is a multiple of \( i_o \ge 1 \) and \( A_i = I \) otherwise. In this case, the arguments of~\cite{Nedic17} apply. This corresponds to a deterministic variant of Federated Averaging, where agents interlace multiple local updates with any round of communications. Of course, variants of these constructions are possible, and we refer the reader to~\iftoggle{short}{\cite{Zhao15, Nedic17}}{\cite{Zhao15, Nedic17, Kairouz21}} for details.

\vspace{-8mm}\textcolor{black}{\section{Conclusion}
The ever increasing need for processing signals and information available at dispersed locations has led to broad research efforts across a number of communities in the past two decades. We have presented a unified view on algorithms for distributed inference and learning through the lens of stochastic primal and primal-dual optimization, and have surveyed some common themes in performance, such as the impact of learning rate, network topology, and the benefit of cooperation. The key take-away of these studies is that in most cases, distributed solutions with appropriately designed cooperation protocols are able to match the performance of centralized, fusion-center based approaches, while offering scalability, robustness to node and link failure, communication efficiency, and no need for the exchange of raw data.}

\smallskip
\noindent\textbf{Stefan Vlaski} is Lecturer at Imperial College London, United Kingdom.

\noindent\textbf{Soummya Kar} is Professor at Carnegie Mellon University, USA.

\noindent\textbf{Ali H. Sayed} is Professor and Dean of Engineering at EPFL, Switzerland.

\noindent\textbf{Jos{\'e} M.~F.~Moura} is the Philip L.~and Marsha Dowd University Professor at CMU, USA.

\vspace*{-.25cm}
\singlespacing
\bibliographystyle{IEEEbib}
{\footnotesize \bibliography{main,CentralBib}}

\end{document}

%% file: custom/operators.tex
\usepackage{amsthm}

\DeclareMathOperator{\cw}{{\scriptstyle\mathcal{W}}}
\DeclareMathOperator{\bcw}{{\boldsymbol{\scriptstyle\mathcal{W}}}}

\DeclareMathOperator{\E}{\mathds{E}}

\DeclareMathOperator{\w}{\boldsymbol{w}}
\DeclareMathOperator{\x}{\boldsymbol{x}}

%% file: custom/environments.tex
\usepackage{amsthm}

\newtheorem{example}{Example}

\newtheorem{condition}{Condition}